\documentclass[12pt]{article}
\usepackage{amssymb} 
\usepackage{mathtools} 
\usepackage{amsthm} 
\usepackage[pdftex]{graphicx}
\usepackage[pdftex]{color}
\usepackage{natbib}
\usepackage{subcaption}
\usepackage[shortlabels]{enumitem}
\usepackage{placeins}
\usepackage{geometry}
\begin{document}

\newcommand{\rar}{\rightarrow}
\newcommand{\lrar}{\longrightarrow}
\newcommand{\oli}{\overline}
\makeatletter
\def\blfootnote{\xdef\@thefnmark{}\@footnotetext}
\makeatother

\newtheorem{Def}{Definition}[section]
\newtheorem{Rm}[Def]{Remark}
\newtheorem{Ex}[Def]{Example}

\def\spacingset#1{\renewcommand{\baselinestretch}%
{#1}\small\normalsize} \spacingset{1}


\title{\bf Torus Principal Component Analysis with an Application to RNA Structures}
\author{Benjamin Eltzner$^{1,*}$, Stephan Huckemann$^{1,*}$, Kanti V. Mardia$^2$
  }
\date{\today}

\blfootnote{$^{1}$ Felix-Bernstein-Institute for Mathematical Statistics in the Biosciences, Georg-August-University G\"{o}ttingen}
\blfootnote{$^{2}$ Department of Statistics, University of Oxford and Department of Statistics, University of Leeds}
\maketitle
\blfootnote{$^*$ The authors gratefully acknowledge DFG CRC 755 and the Niedersachsen Vorab of the Volkswagen Foundation}

\bigskip
\begin{abstract}
  There are several cutting edge applications needing PCA methods for data on tori and we propose a novel torus-PCA method with important properties that can be generally applied. There are two existing general methods: tangent space PCA and geodesic PCA. However, unlike tangent space PCA, our torus-PCA honors the cyclic topology of the data space whereas, unlike geodesic PCA, our torus-PCA produces a variety of non-winding, non-dense descriptors. This is achieved by deforming tori into spheres and then using a variant of the recently developed principle nested spheres analysis. This PCA analysis involves a step of small sphere fitting and we provide an improved test to avoid overfitting. However, deforming tori into spheres creates singularities. We introduce a data-adaptive pre-clustering technique to keep the singularities away from the data. For the frequently encountered case that the residual variance around the PCA main component is small, we use a post-mode hunting technique for more fine-grained clustering. Thus in general, there are three successive interrelated key steps of torus-PCA in practice: pre-clustering, deformation, and post-mode hunting. We illustrate our method with two recently studied RNA structure (tori) data sets: one is a small RNA data set which is established as the benchmark for PCA and we validate our method through this data. Another is a large RNA data set (containing the small RNA data set) for which we show that our method provides interpretable principal components as well as giving further insight into its structure.
\end{abstract}

\noindent%
{\it Keywords:} Statistics on manifolds, tori deformation, directional statistics, dimension reduction, dihedral angles, angular clustering, fitting small spheres, principle nested spheres analysis.

\section{Introduction}

With the rise of the internet, large biomolecule databases, see, for example, \cite{RCSB-PDB}, have become publicly available and further the increased computational power has led to a surge in statistical evaluation. In particular, there are cutting edge applications in structural bioinformatics needing PCA methods for data on a torus, for examples, for RNA structural data (see, for example, \cite{Sargsyan2012}) and for protein structural data (see, for example, \cite{Altis2008}). However dimension reduction on non-Euclidean manifolds with PCA-like methods has been a challenging task.

\begin{figure}[ht!]
  \centering
  \subcaptionbox{Flat torus as square in $\mathbb{R}^2$ with edges identified.\label{torus-winding-geodesica}}[0.45\textwidth]{\includegraphics[width=0.45\textwidth, clip=true, trim=2cm 0 2cm 0]{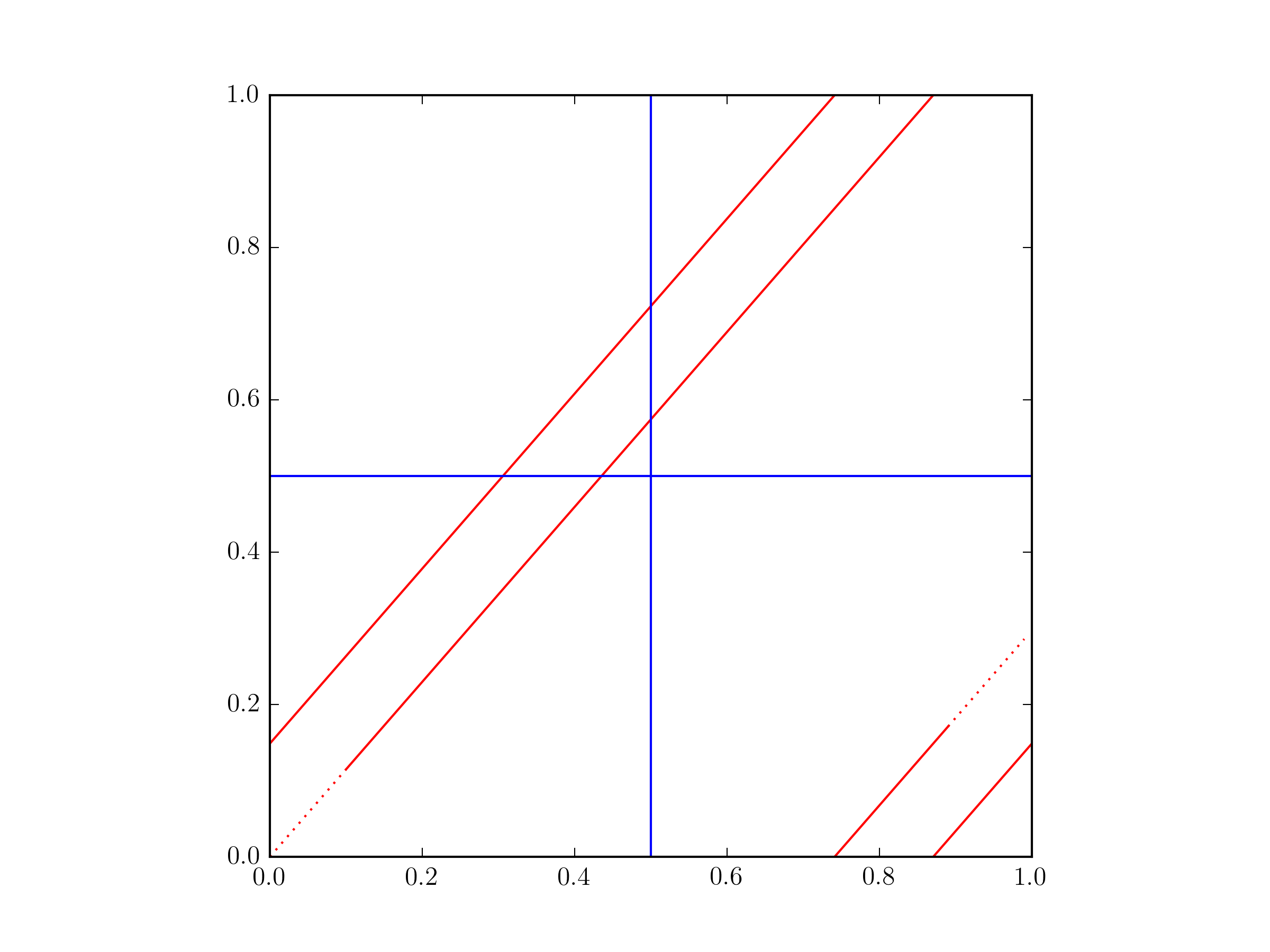}}
  \hspace*{0.05\textwidth}
  \subcaptionbox{Curved torus embedded in $\mathbb{R}^3$.\label{torus-winding-geodesicb}}[0.45\textwidth]{\includegraphics[width=0.45\textwidth, clip=true, trim=1cm 0.5cm 0.5cm 0.5cm]{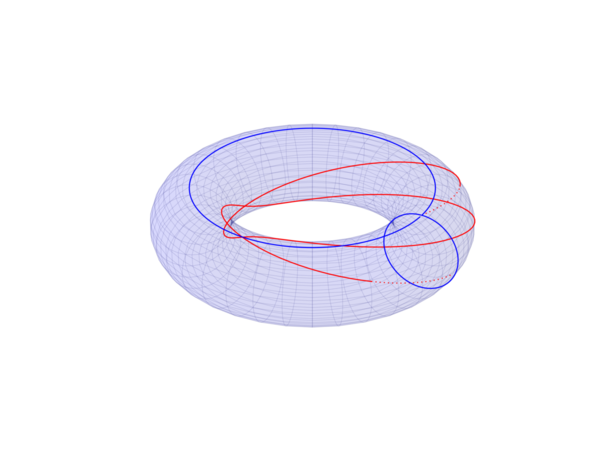}}
  \caption{\it Flat (\ref{torus-winding-geodesica}) and curved (\ref{torus-winding-geodesicb}) torus representation. Except for horizontal and vertical geodesics (blue) in \ref{torus-winding-geodesica} all other geodesics wind around. All geodesics (red) with an irrational slope in \ref{torus-winding-geodesica} are dense.}
  \label{torus-winding-geodesic}
\end{figure}

There are two usually successful categories of methods which have been developed in the last decades: tangent space PCA (extrinsic approach, see, for example, \cite{Fletcher2004,BoisvertPennecLabelleCherietAyache2006,ArsignyCommowickPennecAyache2006}), and geodesic PCA (intrinsic approach, see, for example, \cite{Huckemann2006,Sommer2013}). However, for the very simple non-Euclidean case of the flat and compact space of a torus (a direct product space of two or more angles), these approaches are not adequate. Namely, tangent space PCA fails to take into account the periodicity of the torus and, even worse, geodesic PCA is completely inapplicable because almost all geodesics are densely winding around as seen in Figure \ref{torus-winding-geodesic}.

In this paper we propose the novel tool of torus-PCA (T-PCA), which does not suffer from these defects. This is achieved by deforming tori into spheres and then using a variant of the recently developed principle nested spheres analysis (PNS) of \cite{Jung2012}. This PNS analysis involves a step of small sphere fitting and we provide an improved test to avoid overfitting. However, deforming the geometry of the torus into that of a sphere -- locally glued to itself (to honor periodicity) -- creates singularities. We introduce a data-adaptive pre-clustering technique to keep the singularities away from the data. We then apply the torus deformation to clusters separately. Further, mode hunting is utilized to deal with the case of large variance explained by the 1D PC. To sum up, our full T-PCA algorithm (Section \ref{sec:flow_chart}) consists of three successive steps: pre-clustering, deformed torus PNS (DT-PNS = torus deformation with altered PNS) and post-clustering,

We illustrate the power of our method, using two important RNA data sets. Indeed, the data sets stem from analyses of RNA folding which is believed to be a centerpiece in within-cell communication, see, for example, \cite{Chapman1998,Chakrabarti2011,Brewer2013}. The folding structure is usually described by \emph{dihedral angles} between neighboring planes, each spanned by three adjacent atoms, similar to pages of an opened book (see Appendix for an illustration). Each nucleic base corresponds to a backbone segment described by $6$ angles and one angle for the base, giving a total of $7$ angles. Understanding the distribution of these $7$ angles over large samples of RNA strands is an intricate problem that has drawn some attention, e. g. \cite{Murray2003,Schneider2004,Wadley2007,Richardson2008,Frellsen2009}.

Simulation studies are frequently used to model and understand interactions of RNA strands with proteins occurring in cells, see \cite{Hermann1999,Magee2005,Zhao2006,Estarellas2015}. As the computational complexity of full molecular dynamics simulations is very high, there is a large demand for concisely reduced models obtained from investigations of the RNA conformation space. One way of reducing complexity consists in representing the data in a lower dimensional subspace as done by PCA. Another power of PCA lies in providing continuity to a discretely sampled conformational space as in \cite{Frellsen2009}. For lack of satisfactory torus PCA-methods, previous studies of RNA residue geometry have made use of the two pseudo-torsion angles $\eta$ and $\theta$ (see Figure \ref{rna_backbone_schem}), to accomplish a lower dimensional data representation. These $\eta$--$\theta$ plots (see Figures \ref{geoPCA_labelsa} and \ref{clusters_overlapa}), projecting a two-dimensional torus onto the plane, are called Ramachandran plots for example by \cite{Duarte1998}.

Some torus-specific PCA approaches have been developed apart from tangent space PCA and geodesic PCA. Using wrapped normals, \cite{KentMardia2009} circumvent the problem of winding geodesics and provide for an intrinsic parametric model with the same number of degrees of freedom as classical PCA, which, as discussed in \cite{HuckemannEltzner2015}, is less than the number of degrees of freedom for our type of approach. The PCA used by \cite{Altis2008} is a particular case of \cite{KentMardia2009}. Allowing geodesics only that wind around at most once, as proposed by \cite{KentMardia2015}, further reduces the degrees of freedom.

It seems that \cite{Sargsyan2012} have been the first and only to treat toroidal data describing RNA structures in a spherical geometry. In their construction, they halved the corresponding seven torus angles and treated these as polar angles from a seven-dimensional sphere, thus taking only a very first step towards T-PCA. On this seven-dimensional sphere they investigated a test data set consisting of $190$ residues. However, \cite{Sargsyan2012} did neither discuss nor exploit the drastic change of geometry and only applied geodesic PCA, see \cite{Huckemann2006}, maximizing projected variance and not minimizing residual variance. Incidentally, some pitfalls of using projected variance for compact manifolds have been pointed out in \cite{Huckemann2010}.

In our applications, first we use the \textit{small RNA data set} used by \cite{Sargsyan2012} as a benchmark for our T-PCA method. We find that T-PCA retrieves the underlying clusters in an effective way. Then we analyze a classical data set consisting of 8301 residues, subsequently called the \textit{large RNA data set}, which was carefully selected for high experimental X-ray precision ($0.3$ nanometers) by \cite{Duarte1998,Wadley2007} and analyzed by them and others, for example \cite{Murray2003,Richardson2008}. The small RNA data set is a subset of the large RNA data set consisting of neighborhoods of three known cluster centers in the $\eta$--$\theta$-plot (as in Figure \ref{geoPCA_labelsa}). We compare our method to tangent space PCA and show that T-PCA captures much more data variation in low dimensional subspaces, explaining at least $80 \%$ of the data variance in the one-dimensional representation whereas in contrast tangent space PCA requires at least two dimensions. Beyond two rather well known clusters we identify a new cluster which has not been found previously.

The plan of the paper is as follows: In Section \ref{sec:DT-PNS} we introduce DT-PNS, which is the center piece of our methodology. After reviewing the auxiliary clustering methods in Section \ref{sec:clustering}, we present our torus PCA algorithm in Section \ref{sec:flow_chart}. In Section \ref{sec:appl_rna} we apply our method to the small and large RNA data sets and review the results. The paper ends with a discussion. A brief overview of our abbreviations and technical terms used throughout this paper is given in the Appendix.

\section{Deformed Torus PNS}\label{sec:DT-PNS}

\subsection{Torus Deformation Schemes}\label{sec:sausage_surg}

Let $T^{D}=(\mathbb{S}^{1})^{\times D}$ be the $D$-dimensional unit torus and $\mathbb{S}^D=\{x\in \mathbb R^{D+1}:\|x\|=1\}$ the $D$-dimensional unit sphere, $D\in \mathbb N$. The definition of the data-adaptive deformation mapping $P: T^D \lrar \mathbb{S}^D$ defined in this section is based on a comparison of Riemannian squared line elements. If $\psi_k\in \mathbb{S}^1=[0,2\pi]/\sim$ ($k=1,\ldots, D$) where $\sim$ denotes the usual identification of $0$ with $2\pi$, the squared line element of $T^D$ is given by the Euclidean 
\begin{align*}
  ds^2 &= \sum_{k=1}^{D} d\psi_{k}^2\,.
\end{align*}  
For $\mathbb{S}^D$, in polar coordinates $\phi_k \in [0, \pi]$ for $k=1,\ldots, D-1$ and $\phi_D \in [0, 2\pi]/\sim$, whose relation to embedding space coordinates $x_k$ is elaborated in the Appendix, the spherical squared line element is given by
\begin{align}
  ds^2 &= d\phi_{1}^2 + \sum_{k=2}^{D} \left( \prod_{j=1}^{k-1} \sin^2 \phi_{j} \right) d\phi_{k}^2 \label{trafo_line_elem}\,.
\end{align}
In fact, this squared line element is not defined for the full sphere but only for $\phi_k \in (0, \pi)$ ($k=1,\ldots, D-1$). The singularities at $\phi_k = 0, \pi$ will account for singularities of $P$ which form a subtorus  of dimension $D-2$ (or a union of self-glued subtori). Because in (\ref{trafo_line_elem}), $d\phi^2_1$ comes with the factor $1$, no deformation at all occurs for $\phi_1$, i.e. this angle corresponds to spherical distances without distortion. In the summation for $k=2$ we have a factor $\sin^2\phi_1$ of $d\phi^2_2$, which shows how the angle $\phi_1$ distorts the angle $\phi_2$ and finally the deformation factor $\prod_{j=1}^{D-1} \sin^2 \phi_{j}$ of $d\phi^2_D$ reflects the distortions of $\phi_D$ by all other angles. For this reason, in the following, we will refer to $\phi_{D}$ as the \textit{innermost angle} and to $\phi_{1}$ as the \textit{outermost angle}.

\begin{Rm}\label{Distortion:rmk}
At this point note that near the equatorial great circle given by $\phi_k = \frac{\pi}{2}$ ($k=1,\ldots,D-1$) this squared line element is nearly Euclidean. Distortions occur whenever leaving the equatorial great circle. More precisely, distortions are higher when angles $\phi_k$ with low values of the index $k$ are close to zero, than when angles $\phi_k$ with high values of the index $k$ are close to zero.
\end{Rm}

\begin{Def}[Torus to Sphere Deformation]
With a data-driven permutation $p$ of $\{1, \dots, D\}$, data-driven central angles $\mu_k$ ($k=1,\ldots, D$) and data-driven scalings $\alpha_k$, all of which are described below, set
\begin{align} \label{deformation-def}
  \phi_k = \frac{\pi}{2} + \alpha_{p(k)} (\psi_{p(k)} - \mu_{p(k)}),\quad k=1,\ldots,D\,
\end{align}
where $p(k)$ is the index $k$ permuted by $p$ and the difference $(\psi_{p(k)} - \mu_{p(k)})$ is taken modulo $2\pi$ such that it is in the range $(-\pi, \pi]$. 
\end{Def}

\paragraph{In general, the scalings} are restricted to the choices $\alpha_{k'} = 1/2$ and $\alpha_{k'} = 1$, $k'=p(k)$. If all of the $k'$-th torus angles of the data are within an interval of length $\pi$, choose $\alpha_{k'}=1$ ($k'=1,\ldots,D-1$) leading to \emph{unscaled} (U) angles. Else choose $\alpha_{k'} = 1/2$  ($k'=1,\ldots,D-1$) leading to \emph{halved} (H) angles. The innermost angle will always remain unscaled, $\alpha_D=1$. In practical situations, the torus data are often spread out over more than half circles for several angles. Then we choose (H) angles. In rare cases where data is concentrated we can choose (U) angles.

\paragraph{The central angles} ($\mu_k$) will be chosen such that data points come to lie near the equatorial great circle and omit the singularities. Two plausible choices are:
\begin{enumerate}
  \item[(i)] with the circular intrinsic mean $\oli{\psi}_{k, \text{intr}}$ (we use the fast algorithm from \cite{HoHu2014}), set $\mu_k = \oli{\psi}_{k, \text{intr}}$ to obtain \textit{mean centered} (MC) data
  \item[(ii)] with $\psi_{k, \text{gap}}$, the center of the largest gap between neighboring $\psi_k$ values of data points and $\psi^*_{k, \text{gap}}$ its antipodal point, define $\mu_k = \psi^*_{k, \text{gap}}$ to obtain \textit{gap (antipode) centered} (GC) data.
\end{enumerate}
MC data has the merit that the torus mean of the data is mapped to the equatorial great circle and thus, in that sense, deformation of the data is minimized. For a strongly skewed data distribution, spread out over a half circle, halved GC data will still be confined to a $\pi/2$ neighborhood of the equator while halved MC data will touch the singularities, leading to high distortion there. For data sets with outliers, GC centering may not be robust, making MC more favorable.

\paragraph{The choice of the permutation} ($p$) is driven by analyses of the \textit{data spread}
\begin{align}
  \sigma_k^2 = \sum_{i = 1}^n \limits (\psi_{k,n} - \mu_k)^2
\end{align}
for each angle, where $\psi_{k,i}\in \mathbb{S}^1 $ are the torus data and $n$ is the number of data points on $T^D$. If the angles are ordered by increasing data spread, such that $\sigma_{p(1)}^2$ is minimal and $\sigma_{p(D)}^2$ is maximal, in view of Remark \ref{Distortion:rmk}, the change of distances between data points caused by the deformation factors $\sin^2 \phi_{j}$ in Equation \eqref{trafo_line_elem} is minimized. We call this case \textit{spread inside} (SI), because variation is concentrated on the inner angles of the sphere. The opposite ordering is called \textit{spread outside} (SO). We will restrict consideration to these two options.

Due to periodicity on the torus, $\psi_k=0$ is identified with $\psi_k=2\pi$ for all $k=1,\ldots,D$. In contrast, for all angles $\phi_k=0$, with $k=1,\ldots,D-1$, denotes spherical locations different from $\phi_k=\pi$. For an invariant representation respecting the torus' topology, however, it is necessary to identify these locations accordingly, which results in a \textit{self-gluing} of the $\mathbb{S}^D$ as elaborated in more detail in the Appendix.

\subsection{Linking the Torus' Deformation to PNS}\label{subsec:dt-pca}

For data sets on a torus, we apply a deformation as detailed in Section \ref{sec:sausage_surg}, in particular a data driven choice of scalings ($\alpha_k = 1/2$ if the data in one of the angles except for the innermost is spread out on more than a half circle or else $\alpha_k = 1$) and of centering (MC or GC) is performed.
On the resulting self-glued $\mathbb{S}^D$ we use an alteration of principal nested sphere analysis (PNS) by \cite{Jung2010,Jung2012} for dimension reduction.

The PNS iteration leads to a sequence of small subspheres
\begin{align}
\mathbb{S}^D \supset S^{D-1} \supset \dots \supset S^2 \supset S^1 \supset \{ \mu\} . \label{subspace_sequence}	
\end{align}
The ultimate point $\mu$ is called the \emph{nested mean}. For real data applications, with probability one, the $S^d$ are not great subspheres but proper small subspheres ($d=1,\ldots,D-1$), the radii of which decrease monotone (as discussed further in Section \ref{scn:improvedPNS}). At each reduction step, the residues are given as signed distances: points lying inside the small subsphere receive a positive distance, points lying outside a negative distance.

The PNS algorithm consists of two parts which alternate, namely the fitting of a subsphere $S^{d-1}$ and the projection to this subsphere $\pi_d : S^d \rar S^{d-1}$ ($d=D,D-1,\ldots,1$). If $\mathbb{S}^D$ is glued to itself, in the fitting step as well as in the projection step, distances through the glued part (which may be shorter than the spherical distance) can be used instead of spherical distances only, as in classical PNS. We find from experiments that the fitting procedure when taking into account gluing is numerically badly behaved and tends toward local minima, even more so if we use $\tilde{\delta}$ introduced below in (\ref{delta-tilde:def}). For this reason we alter classical PNS by taking into account the topological identifications only in the projection step. Thus, data fidelity is preserved while the simplified choice of subspace sequences is supported by the resulting good low dimensional description of the data.

\subsection{Comparing Variances}\label{sec:variances}

In Euclidean spaces, PCA variances are additive with monotone decrements leading to a convex variance plot as a property of the metric. In non-Euclidean spaces, this is no longer the case (see the discussion for various definitions of intrinsic variances in \cite{Huckemann2010}). Even worse, comparing variances of different clusters is further complicated by each cluster being analyzed in its own data-adaptive deformed torus. In order to perform a meaningful comparison of variances, we propose the following calculation of residual variances as measures for the quality of the fit. 

Assume a cluster $\mathcal{C}$ and a corresponding adaptive deformation $P_\mathcal{C} : T^D \rar \mathbb{S}^D$. Using the inverse deformation $P_\mathcal{C}^{-1}$ (which is well defined except for the singularities) and the torus metric
\begin{align*}
  \delta : T^D \times T^D \rar \mathbb{R}^{\ge 0} \qquad (p, q) \mapsto \left(\sum_{i = 1}^D \min \big( |p_i - q_i|^2, (2\pi - |p_i - q_i|)^2 \big) \right)^{\frac{1}{2}}
\end{align*}
we define the following function on the sphere
\begin{align}\label{delta-tilde:def}
  \begin{array}{rl}
  \tilde{\delta} : \mathbb{S}^D \times \mathbb{S}^D \rar \mathbb{R}^{\ge 0} \qquad (x,y) \mapsto \delta\big(P_\mathcal{C}^{-1}(x),P_\mathcal{C}^{-1}(y)\big)\, .
  \end{array}
\end{align}
This is a metric when we take into account the topological identifications. Recall that PNS yields a sequence of subspaces $\mathbb{S}^D \supset S^{D-1} \supset \dots \supset S^1 \supset \{\mu\}$ with projections $\pi_d : S^{d+1} \rar S^d \subset S^{d+1}$, $\pi_0 : S^1 \rar \{\mu\}$. From these we define the iterated projections
\begin{align*}
  \Pi_d = \pi_d \circ \pi_{d+1} \circ \dots \circ \pi_{D-1}
\end{align*}
and finally the residual variances (variance not explained by $S^d$)
\begin{align*}
  V_{\mathcal{C}, P_\mathcal{C}, d} = \sum_{q \in \mathcal{C}} \limits \tilde{\delta}^2 (q, \Pi_d(q))\,.
\end{align*}
Due to nestedness, these sequences are non-increasing with $d$. However, the decrements $V_{\mathcal{C}, P_\mathcal{C}, d-1} - V_{\mathcal{C}, P_\mathcal{C}, {d}}$ ($d=1,\ldots,D$) are not necessarily non-increasing, so the resulting curve in the variance plot need not be convex as seen in Figure \ref{torus_scree} (discussed in Section \ref{sec:large_data}). This is in contrast to the Euclidean case, where the plot is always convex because decrements correspond to the non-increasingly ordered eigenvalues of the corresponding covariance matrix. In order to honor differences of densities over different clusters, we normalize all residual variances by dividing by the common scale given by the {\it total variance}
\begin{align}
V_0 := \min_{p \in T^D} \sum_{q\in {\mathcal{Z}}} \tilde{\delta}^2(q,p)
\end{align}
over the full data set $\mathcal{Z}$. If we would individually normalize the residual variances of a cluster $\mathcal{C}$ by its total variance or by $V_{\mathcal{C}, P_\mathcal{C}, 0}$ then a concentrated and isotropic cluster would yield a nearly linear residual variance plot, suggesting that the data may be high dimensional. Normalizing by the common scale, however, still yields nearly a line, now starting well below $100 \%$ at the zero-dimensional approximation, more realistically suggesting that the data is zero-dimensional (see Figures \ref{torus_scree} and \ref{clusters_scree}).

\subsection{Improved PNS: Avoiding Overfitting}\label{scn:improvedPNS}

In the PNS algorithm a cluster of points concentrated around a single center may still be best fitted by a very small subsphere. As this obvious overfitting is undesirable, \cite{Jung2011,Jung2012} would rather fit a great subsphere in such cases and give tests for this purpose. We propose an improved test based on  geometrically better hypotheses and a likelihood ratio, detailed in the Appendix.

We carried out a simulation study to compare the tests. Test data conforming to the null hypothesis (a cluster leading to a great sphere fit) have been generated, by simulating isotropically normal distributed points in a tangent plane with standard deviations $\sigma$ uniform in $[0.1, 0.45]$ truncated to  $2\sigma$ and projecting these points orthogonally to the sphere. Test data for the  alternative (small sphere) has a uniform angular distribution and a non-centered normal radial distribution with means $\mu_r$ uniform in  $[0.1, 0.5]$ and $\sigma$ uniform in $[0.01\mu_r, 0.5\mu_r]$. The resulting distribution is then truncated to the unit circle. The results are displayed in Table \ref{test_comparison}. In effect of modeling specific hypotheses (among others a ring with a central cluster as detailed in the Appendix), the errors of the first kind of the tests by \cite{Jung2011,Jung2012} are unacceptably high, whereas for our test, these are not as high. Our test features approximately the same order for the errors of the second kind, whereas for \cite{Jung2011, Jung2012}, they are much smaller. Our test is clearly an improvement, approximately giving an equal error rate. However, its fine tuning and robust extension to more general data models warrants further investigation beyond the scope of this paper.

\begin{table}[!ht]
  \caption{\textit{Errors of three test; of the first kind (falsely rejecting the null hypothesis of a great sphere) and of the second kind (falsely accepting the alternative of a small sphere) in a simulation with $1000$ test clusters.}}

  \begin{center} {
    \begin{tabular}{|c|c|c|c|c|c|c|}
      \hline
                   & \multicolumn{2}{|c|}{\rule{0pt}{2.6ex} our test}   & \multicolumn{2}{|c|}{\rule{0pt}{2.6ex} \cite{Jung2011}} & \multicolumn{2}{|c|}{\rule{0pt}{2.6ex}  \cite{Jung2012}}\\ \hline \vspace*{0.5\baselineskip}
      Cluster size & 1st kind & 2nd kind & 1st kind & 2nd kind & 1st kind & 2nd kind\\ \hline \vspace*{0.5\baselineskip}
      $30$         & $18.6 \%$  & $21.3 \%$   & $93.8 \%$  &  $1.0 \%$ &  $52.3 \%$  & $9.3 \%$\\ \vspace*{0.5\baselineskip}
      $100$        &  $8.9 \%$  & $18.0 \%$   & $84.8 \%$  &  $3.3 \%$ &  $71.5 \%$  & $5.1 \%$\\ \vspace*{0.5\baselineskip}
      $300$        & $10.9 \%$  & $12.3 \%$   & $80.6 \%$  &  $2.6 \%$ &  $95.7 \%$  & $1.6 \%$\\ \vspace*{0.5\baselineskip}
      $1000$       & $23.6 \%$  & $10.3 \%$   & $84.3 \%$  &  $4.1 \%$ &  $100.0 \%$ & $0.6 \%$\\ \hline
    \end{tabular}
  }\end{center}
  \label{test_comparison}
\end{table}

\section{Pre- and Post-Clustering}\label{sec:clustering}

\subsection{Single Linkage Pre-Clustering}\label{sec:pre-clustering}

\begin{figure}[ht!]
   \centering
   \begin{minipage}{0.4\textwidth}
   \includegraphics[width=1\textwidth, clip=true, trim=0.9cm 0.5cm 0.8cm 0.5cm]{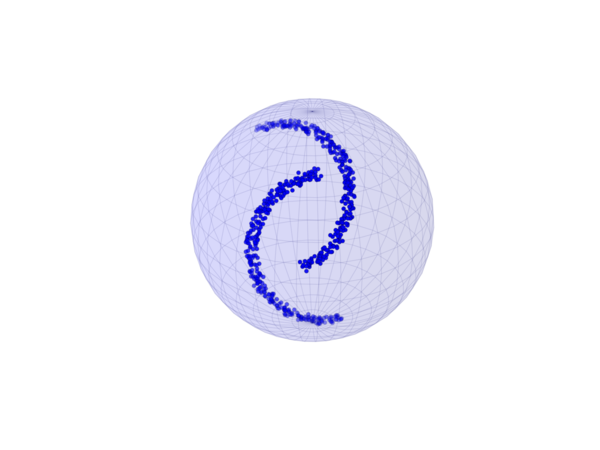}
   \end{minipage}
    \begin{minipage}{0.05\textwidth}\hfill \end{minipage}
  \begin{minipage}{0.4\textwidth}
   \caption{\it An example of two entangled half circles.\label{toy-example:fig}}
   \end{minipage}
   
\end{figure}

We show in Figure \ref{toy-example:fig} a toy example on $T^2$ which highlights two clusters, each of curved one-dimensional data, but entangled. Without treating each cluster separately, DT-PNS fails to discover the one-dimensional structures and suggests a two-dimensional approximation of the joint data. Hence, we need a clustering method that specifically uncovers entangled curved clusters and allows to separate them. It turns out that hierarchical single linkage clustering (also called ``nearest neighbor clustering''), e.g. \cite{MardiaKentBibby1979}, is a suitable method for this task. 

Single linkage clustering joins nearest neighbors recursively to a binary tree structure which is pruned by branch cuts. Each proper node of the tree carries a value denoting the minimal distance between leaves of its two branches, such that node values increase when approaching the root. 

Determining suitable branch cuts for such a cluster tree is a delicate issue that has received attention recently. Especially, methods for data adaptive branch cutting instead of cutting at a fixed level have been proposed, e.g. by \cite{Langfelder2008,Obulkasim2015}. We have designed a similar data adaptive branch cutting recursive procedure detailed in the Appendix. This returns clusters of decreasing density, populating surprisingly low dimensional subspaces of the data space, often entangled as sketched in Figure \ref{toy-example:fig}.

\subsection{Post-Clustering: Mode Hunting}\label{sec:mode-hunting}
As a final step, we analyze the one-dimensional projection of the data using the multiscale method described by \cite{Duembgen2008} for mode hunting. Although this method was originally defined for the real line, its numerical implementation for circular data is even simpler. Since modes are separated by minima, we use this method to identify regions in which minima are located with a certain confidence level. (Throughout the applications, we use a fixed confidence level of $95\%$.) For circular data, we use a wrapped Gaussian kernel smoother for the one-dimensional projections of the points. For every region with minimal smoothed density we increase the kernel's width until there is exactly one minimum of the smoothed distribution left. Here, we separate the modes.

\section{Torus PCA: A Brief Overview}\label{sec:flow_chart}

To a given data set $\cal{Z}$ we first apply DT-PNS as described in Section \ref{sec:DT-PNS} for all deformations (centering with MC or GC and permuting via SI or SO). If for none of these deformations the residual variance to the penultimate small sphere (actually a 1D small circle) is below a threshold of $20\%$, pre-clustering as described in Section \ref{sec:pre-clustering} is performed and DT-PNS as above is applied again for each cluster until the threshold is reached. If none of the deformations achieve a residual variance below $20\%$, the cluster is declared final; otherwise, we apply mode hunting as detailed in Section \ref{sec:mode-hunting}. If mode hunting detects new sub-clusters, these are added to the list of clusters and DT-PNS is performed again on each new sub-cluster. A flow chart for the T-PCA algorithm is included in the Appendix.

\section{Application to RNA Structure}\label{sec:appl_rna}

\begin{figure}[ht!]
   \centering
   \subcaptionbox{\it 3D structure of an RNA residue.\label{rna_backbone_3d}}[0.45\textwidth]{\includegraphics[width=0.45\textwidth]{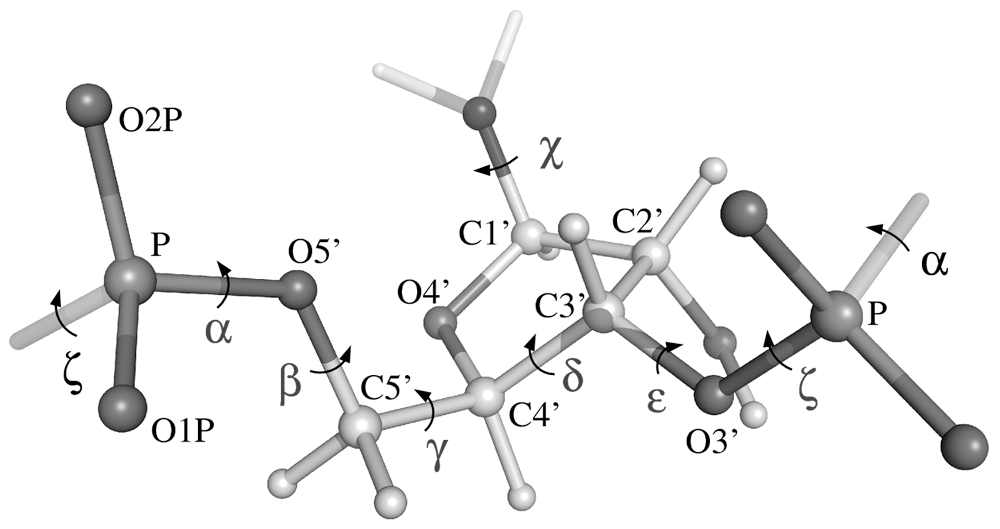}}
   \subcaptionbox{\it 2D scheme of an RNA residue.\label{rna_backbone_schem}}[0.45\textwidth]{\includegraphics[width=0.45\textwidth]{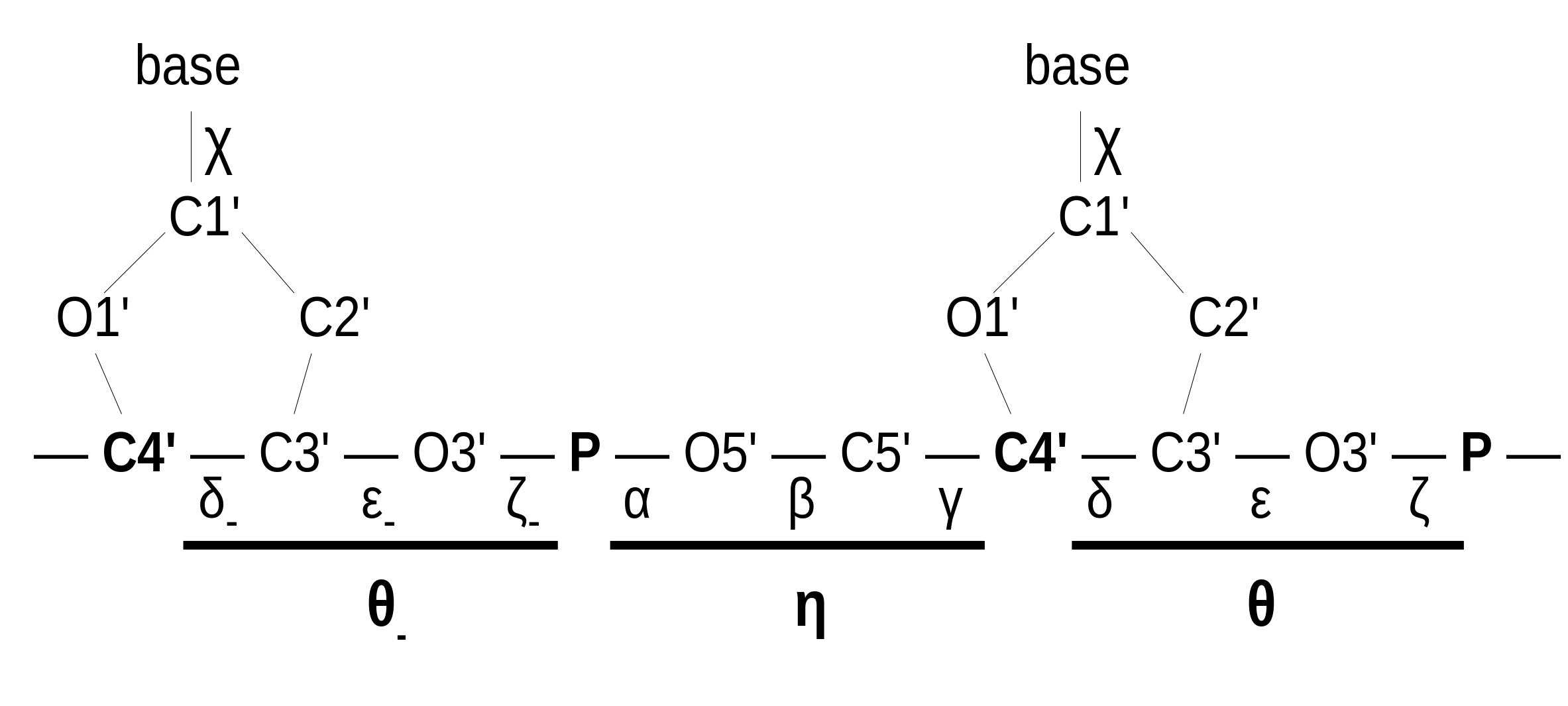}}
   \caption{\it Part of an RNA backbone (Phosphate groups followed by sugars to which a nucleic base it bound). Dihedral angles (Greek letters) are defined by three bonds, the central bond carries the label; pseudo-torsion angles (bold Greek letters) are defined by the pseudo-bonds between bold printed atoms (Figure \ref{rna_backbone_schem}). The precise definition with same canonical atom notation are given in Table \ref{table}. $O$ denotes oxygen, $C$ carbon and $P$ phosphor. The subscript ``$-$'' denotes angles of the neighboring residue. Figure \ref{rna_backbone_3d} is from \cite{Frellsen2009}.}
\end{figure}

For our applications, we need some background on RNA molecules, which is provided in the following. RNA molecules, like DNA, consist of three building blocks, a phosphate group (O3'-P-O5'), a 5-carbon sugar (pentose) and a nucleic base, see Figure \ref{rna_backbone_schem}. The phosphate groups and sugars form the \emph{backbone} in an alternating sequence, where a nucleic base is attached to each sugar, see Figure \ref{rna_backbone_schem}. Each triple is called a \emph{residue} (nucleotide), where, however, for reasons of symmetry, boundaries are not put at phosphate groups beginnings/endings, rather the parts between two consecutive phosphor atoms are considered. The sugar molecules are strictly directed, the C5' to C3' atoms being part of the backbone and the C4' to C1' atoms forming a ring. The nucleic base is attached to the C1' atom, which is furthest from the backbone. Four different standard bases exist, namely Adenine (A), Cytosine (C), Guanine (G) and Uracil (U); while A, C, and G are also common in DNA but  U is replaced by Thymine (T) in DNA. In distinction to the DNA backbone, an oxygen atom is attached to the C2' atom, as displayed in Figure \ref{rna_backbone_3d}. As a result, the C3'-endo sugar pucker (non-planar sugar ring) is energetically preferred and thus by far more common for RNA. For more details we refer the reader  to \cite{Saenger1984}.

In further contrast to DNA, which usually takes a double-stranded helical conformation, RNA is usually single-stranded and the single strand interacts with itself, forming complex shapes. This means that the geometry is much more variable even on the scale of single atoms. Each nucleic base corresponds to a backbone segment described by 6 dihedral angles and one angle for the base, giving a total of 7 angles. Understanding the distribution of these 7 angles over large samples of RNA strands is an intricate problem that has drawn some attention, see \cite{Murray2003,Schneider2004,Wadley2007,Richardson2008,Frellsen2009}. Figure \ref{rna_backbone_3d} details a segment of the RNA backbone with seven angles for each residue giving the 3D folding structure and Table \ref{table} gives the canonical names of the atoms involved in the definition of each angle.

\begin{table}[!ht]
  \caption{\it Seven dihedral angles ($\alpha$, $\beta$, $\gamma$, $\delta$, $\epsilon$, $\zeta$, $\chi$) and two pseudo-torsion angles ($\eta$, $\theta$) in terms of their corresponding four atoms. Figure \ref{rna_backbone_3d} shows the geometry of these atoms. ($N$ denotes nitrogen.)}
  \begin{center}
    \begin{tabular}{|l|c@{$-$}c@{$-$}c@{$-$}c@{  }l|}
      \hline
      $\alpha$   & $O3'$ & $P$    & $O5'$ & $C5'$  &\\ \hline \vspace*{0.5\baselineskip}
      $\beta$    & $P$    & $O5'$ & $C5'$ & $C4'$  &\\ \hline \vspace*{0.5\baselineskip}
      $\gamma$   & $O5'$ & $C5'$ & $C4'$ & $C3'$  &\\ \hline \vspace*{0.5\baselineskip}
      $\delta$   & $C5'$ & $C4'$ & $C3'$ & $O3'$  &\\ \hline \vspace*{0.5\baselineskip}
      $\epsilon$ & $C4'$ & $C3'$ & $O3'$ & $P$     &\\ \hline \vspace*{0.5\baselineskip}
      $\zeta$    & $C3'$ & $O3'$ & $P$    & $O5'$  &\\ \hline \vspace*{0.5\baselineskip}
      $\chi$     & $O4'$ & $C1'$ & $N1$   & $C2$ & for pyrimidine (monocyclic) bases\\
                 & $O4'$ & $C1'$ & $N9$   & $C4$ & for purine (bicyclic) bases \\ \hline \vspace*{0.5\baselineskip}
      $\eta$     & $C4'$ & $P$    & $C4'$ & $P$     &\\ \hline \vspace*{0.5\baselineskip}
      $\theta$   & $P$    & $C4'$ & $P$    & $C4'$  &\\ \hline
    \end{tabular}
  \end{center}
  \label{table}
\end{table}

An approximation of the geometric folding structure on the level of single residues is given by the two \emph{pseudo-torsion angles} $\eta$ and $\theta$ (Figure \ref{rna_backbone_schem} and Table  \ref{table}). These provide at once a two-dimensional visualization (Figure \ref{geoPCA_labelsa}), see e.g.  \cite{Duarte1998,Wadley2007}. Clustering and structure investigation based on the purely backbone torsion angles $\delta_-$, $\epsilon_-$, $\zeta_-$, $\alpha$, $\beta$, $\gamma$ and $\delta$ (see Figure \ref{rna_backbone_schem}) has been performed by \cite{Murray2003,Richardson2008}.


\subsection{The Small RNA Data Set}

\begin{figure}[ht!]
   \centering
   \subcaptionbox{\it $\eta$--$\theta$ plot\label{geoPCA_labelsa}}[0.45\textwidth]{\includegraphics[width=0.45\textwidth, clip=true, trim=2cm 0 2cm 0]{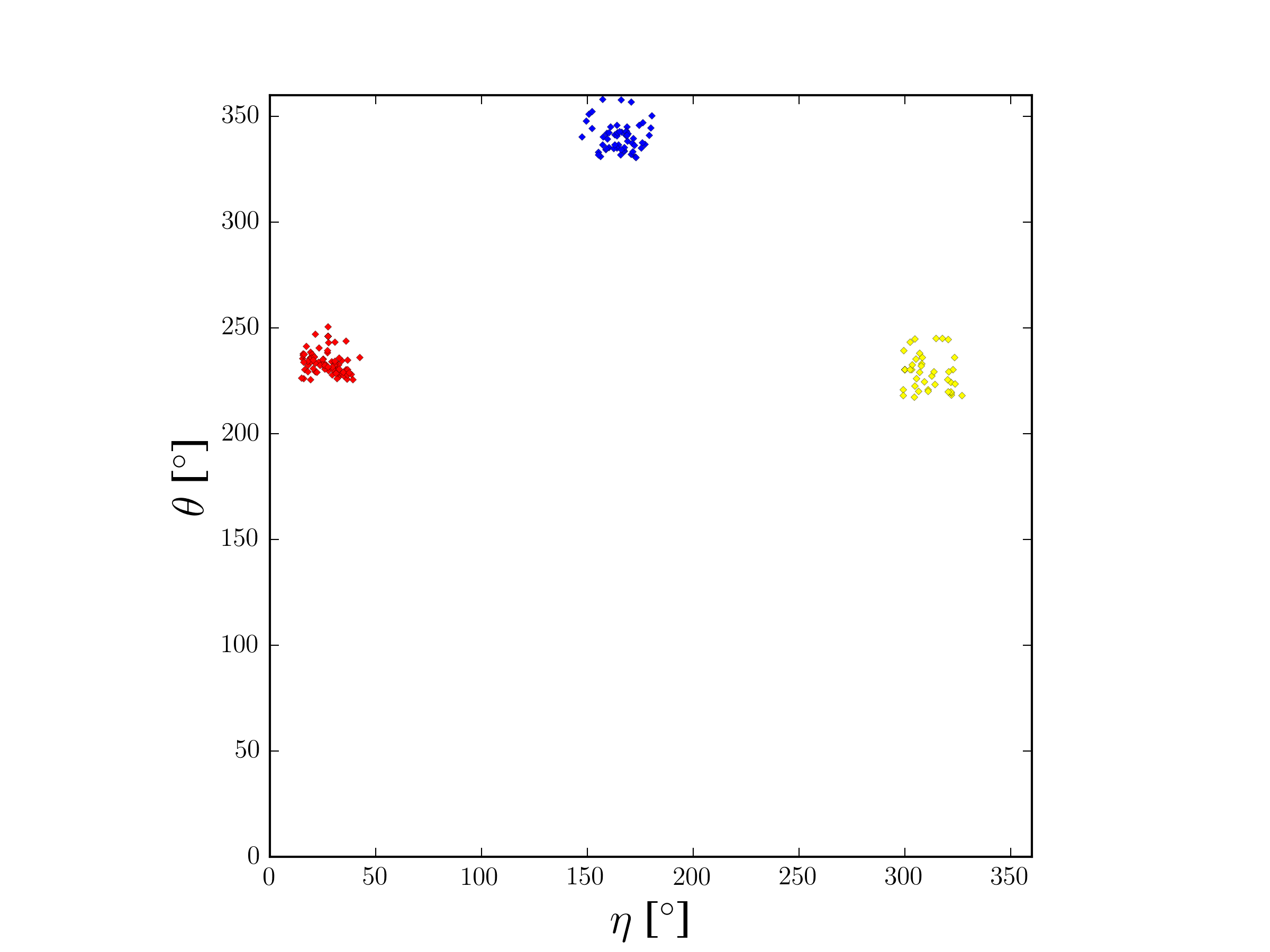}}
   \hspace*{0.05\textwidth}
   \subcaptionbox{\it $\alpha$--$\zeta$ plot\label{geoPCA_labelsb}}[0.45\textwidth]{\includegraphics[width=0.45\textwidth, clip=true, trim=2cm 0 2cm 0]{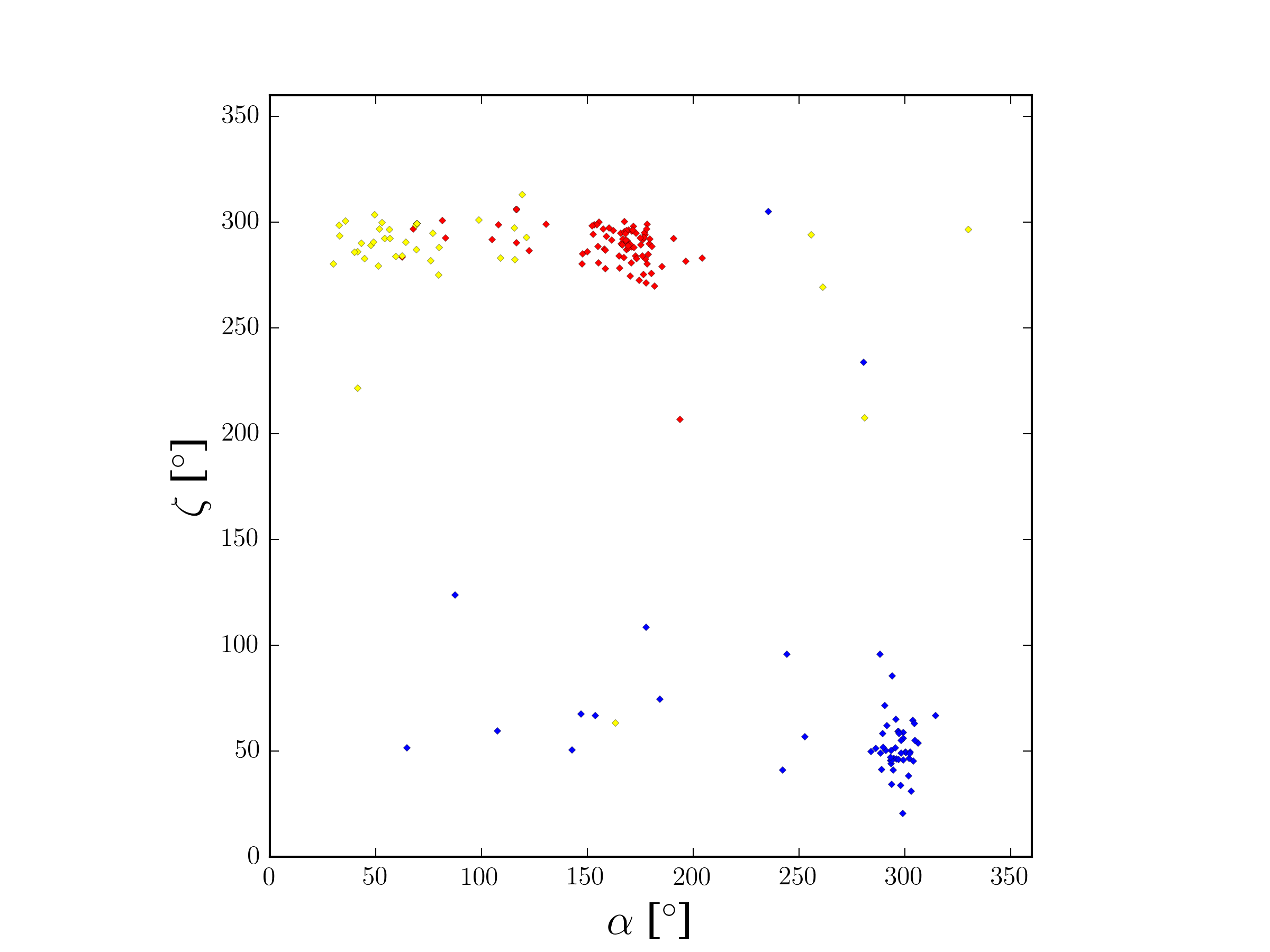}}
   \caption{\it \ref{geoPCA_labelsa}: The small RNA data set of \cite{Sargsyan2012} with their three preselected clusters in the $\eta$--$\theta$ plot. \ref{geoPCA_labelsb}: The small RNA data set plotted in the two most discriminant ($\alpha$, $\zeta$) out of the seven dihedral angles.}
   \label{geoPCA_labels}
\end{figure}

This small RNA data set has been carefully selected by \cite{Sargsyan2012} to validate their method. They took clusters labeled I (blue, $59$ points), II (red, $88$ points) and V (yellow, $43$ points) by \cite{Wadley2007} totaling $190$ data points, which form three clusters in the  $\eta$--$\theta$ plot as shown in Figure  \ref{geoPCA_labelsa}. While clusters I and II correspond to distinct structural elements featuring base stacking, the residues in cluster V belong to a wider variety of structural elements. 

As challenge, however, this distinction cannot be readily seen in the 7D space of all torsion angles. Figure \ref{geoPCA_labelsb} depicts the most discriminant angle pair ($\alpha$, $\zeta$): The yellow cluster is not very concentrated and parts of it are very close to the red cluster, which is twice as big. In fact, upon close inspection, due to periodicity, the red and yellow clusters are also rather close in the $\eta$--$\theta$ plot in Figure \ref{geoPCA_labelsa}.

\begin{figure}[ht!]
  \centering
  \subcaptionbox{\it 2D approximation, SI\label{clusters_spheresa}}[0.45\textwidth]{\includegraphics[width=0.45\textwidth, clip=true, trim=0.5cm 0.2cm 0.5cm 0.2cm]{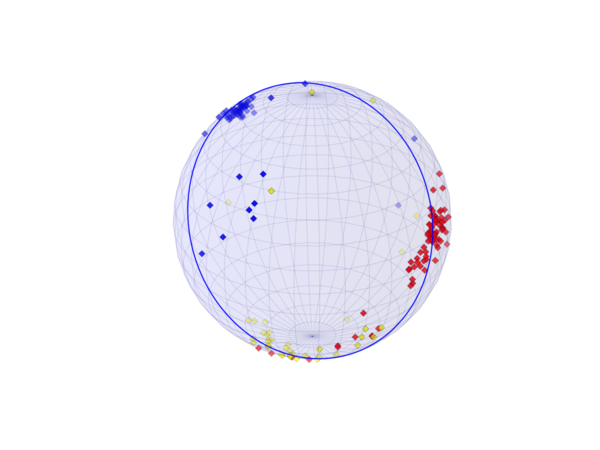}}
  \hspace*{0.05\textwidth}
  \subcaptionbox{\it 2D approximation, SO\label{clusters_spheresb}}[0.45\textwidth]{\includegraphics[width=0.45\textwidth, clip=true, trim=0.5cm 0.2cm 0.5cm 0.2cm]{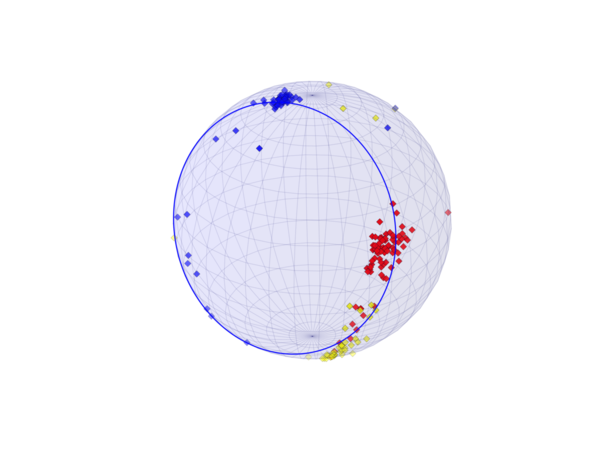}}
  \caption{\it Two-dimensional T-PCA approximation of the small RNA data set with SI (\ref{clusters_spheresa}) and SO (\ref{clusters_spheresb}) ordering. (Colors representing the same clusters as in Figure \ref{geoPCA_labels}).}
  \label{clusters_spheres}
\end{figure}

Without pre-clustering and without post-mode hunting, we have applied T-PCA to all seven angles and depict the two-dimensional representation both for SI (Figure \ref{clusters_spheresa}) and SO (Figure \ref{clusters_spheresb}) ordering in Figure \ref{clusters_spheres}. The data are, in fact, very well approximated by the best fit circle. Using the same coloring for Figure \ref{clusters_spheres} as Figure \ref{geoPCA_labels} shows that the three preselected clusters can be rather well distinguished by eye with slightly better distinction for SO ordering. 

\begin{figure}[ht!]
  \centering
  \subcaptionbox{\it SI $\eta$--$\theta$ plot\label{geoPCA_resultsa}}[0.4\textwidth]{\includegraphics[width=0.4\textwidth, clip=true, trim=2cm 0 2cm 0]{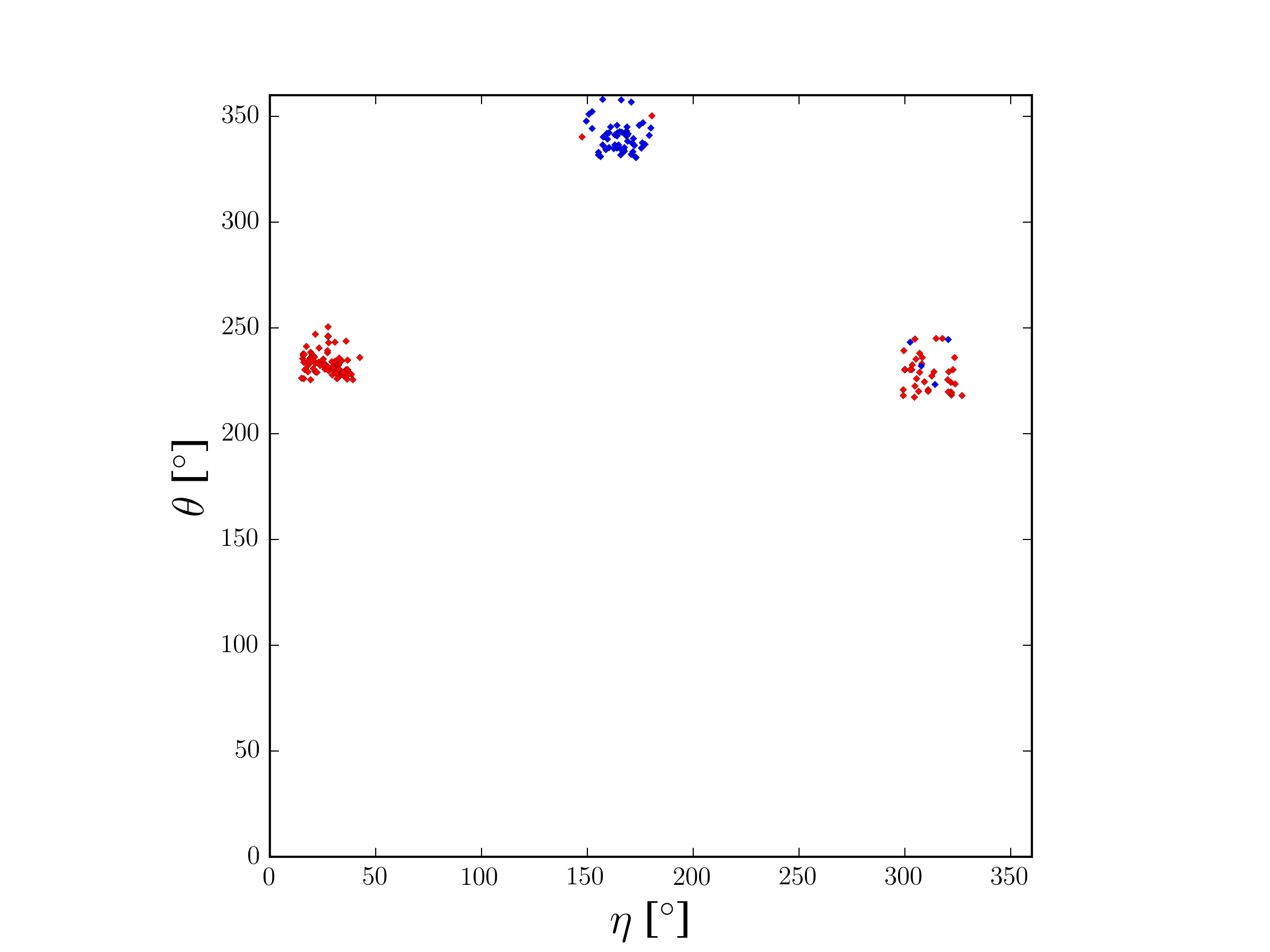}}
  \hspace*{0.05\textwidth}
  \subcaptionbox{\it SO $\eta$--$\theta$ plot\label{geoPCA_resultsb}}[0.4\textwidth]{\includegraphics[width=0.4\textwidth, clip=true, trim=2cm 0 2cm 0]{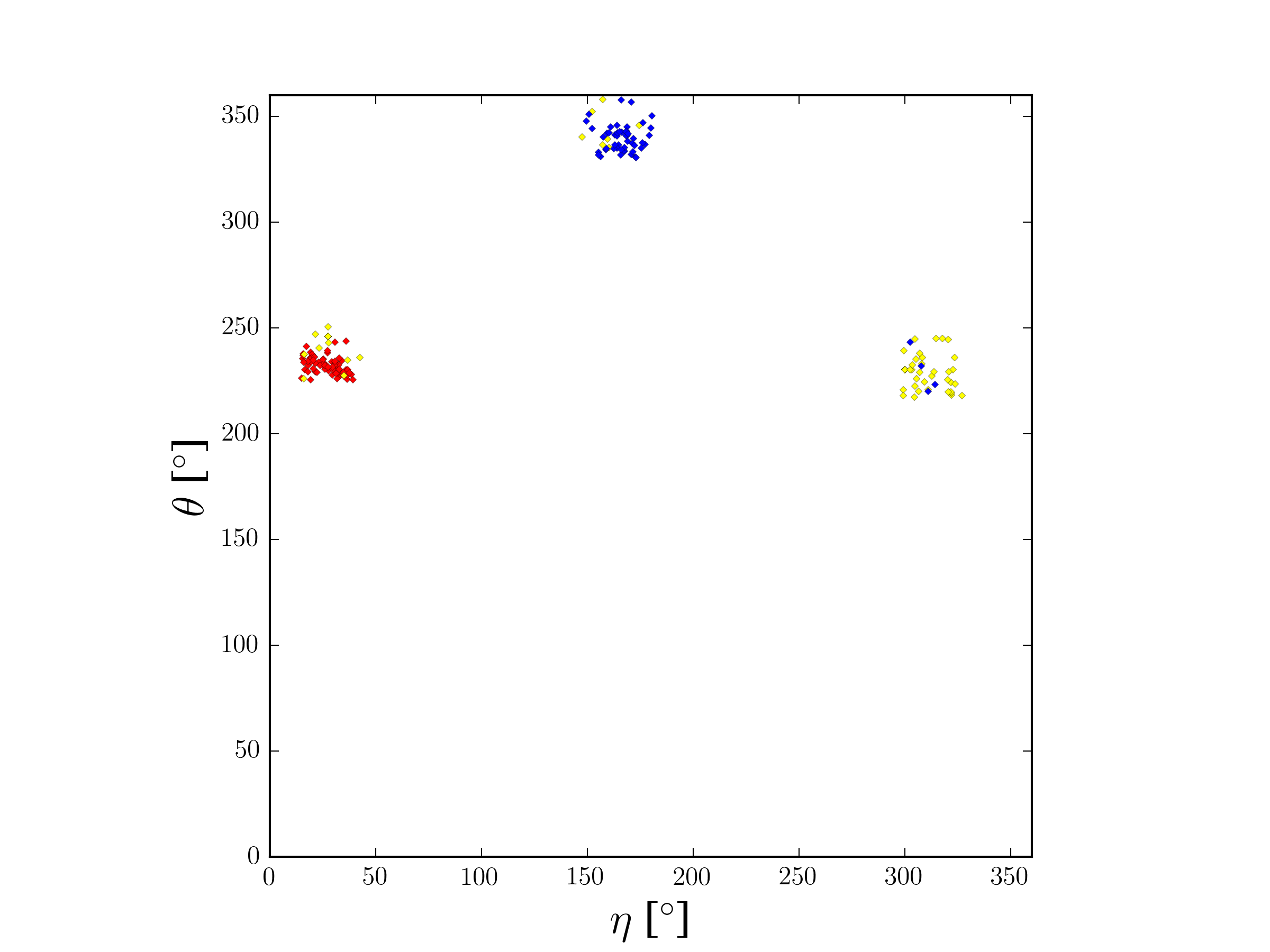}}
  \caption{\it The small RNA data set in pseudo-torsion angles with clusters obtained from T-PCA and labelled with colors red, blue and yellow. \ref{geoPCA_resultsa}: mode hunting and SI. \ref{geoPCA_resultsb}: mode hunting and SO.}
  \label{geoPCA_results}
\end{figure}

When mode-hunting (see Section \ref{sec:mode-hunting}) is applied to the one-dimensional T-PCA representation, for SI-ordering we only find two clusters which correspond roughly to the original blue cluster and the union of the original red and yellow cluster in the $\eta$--$\theta$ plot; recall that the latter are also nearby in the   $\eta$--$\theta$ plot. Using mode hunting with SO ordering the relative distance of the red and yellow cluster is visibly enhanced and we find again three clusters which are only slightly different from the preselected ones. Figure \ref{geoPCA_results} assigns colors to clusters found by T-PCA with mode-hunting and depicts these in the $\eta$--$\theta$ plot. As some outliers occur in the $7$-dimensional representation, we chose mean centered angles.

This result illustrates the power of backward dimension reduction methods going significantly beyond the analysis of \cite{Sargsyan2012}. Not only can the preselected clusters be separated but the data are very accurately approximated by their projection to a circle. Additionally, some points can be identified, which stray so far from the bulk of their designated cluster in the $7$D representation that they are attributed to other clusters by T-PCA (e.g. the blue and yellow points in the left middle in Figure \ref{clusters_spheresa}). Indeed, applying our T-PCA with pre-clustering, we can identify these points as outliers not belonging to any of the three clusters, as elaborated in the Appendix.

\FloatBarrier

\subsection{The Large RNA Data Set}\label{sec:large_data}

\begin{figure}[ht!]
   \centering
   \begin{minipage}{0.4\textwidth}
   \includegraphics[width=1\textwidth]{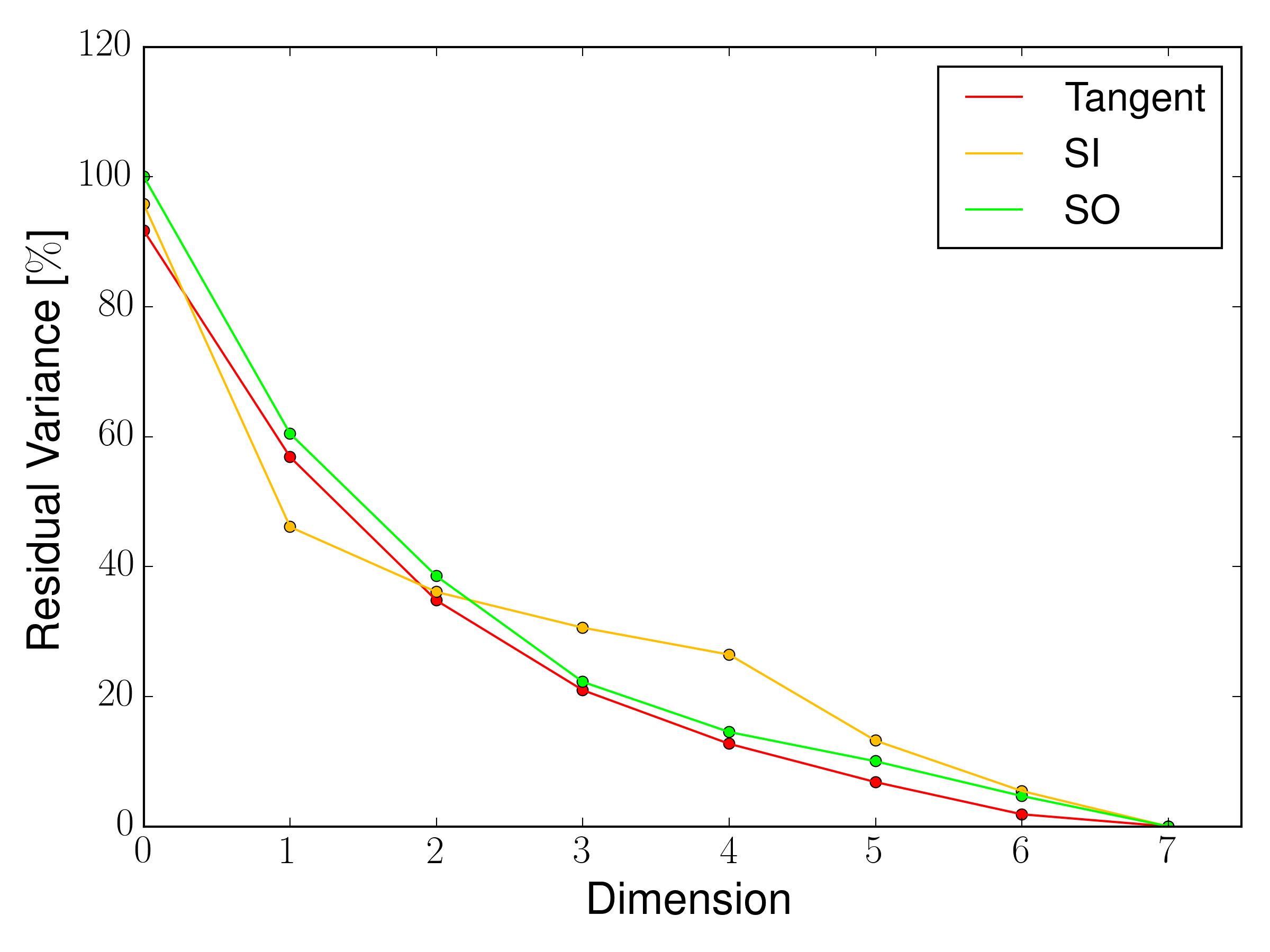}
   \end{minipage}
   \begin{minipage}{0.03\textwidth}\hfill
   \end{minipage}
   \begin{minipage}{0.55\textwidth}
   \caption{\it Large RNA data set: Residual variances for T-PCA (SI and SO without clustering).\label{torus_scree}}
   \end{minipage}
\end{figure}

The large RNA data set consists of 8301 data points. These data spread out widely in almost all seven dihedral angles, so gluing effects must be taken into account for a T-PCA analysis with neither pre- nor post-clustering. To the end of  minimizing the effect of these topological degeneracies we use gap centered angles in this case. The residual variance plot in Figure \ref{torus_scree} of the full data indicates that at least $4$ or $5$ dimensions are necessary to obtain at most $20 \%$ residual variance (residual variance is defined in Section \ref{sec:variances}). To achieve better dimension reduction we thus need to pre-cluster the data.

\subsubsection{PCAs of Clustered Data}

\begin{figure}[ht!]
   \centering
   \subcaptionbox{DT-PCA variances\label{clusters_screea}}[0.45\textwidth]{\includegraphics[width=0.45\textwidth]{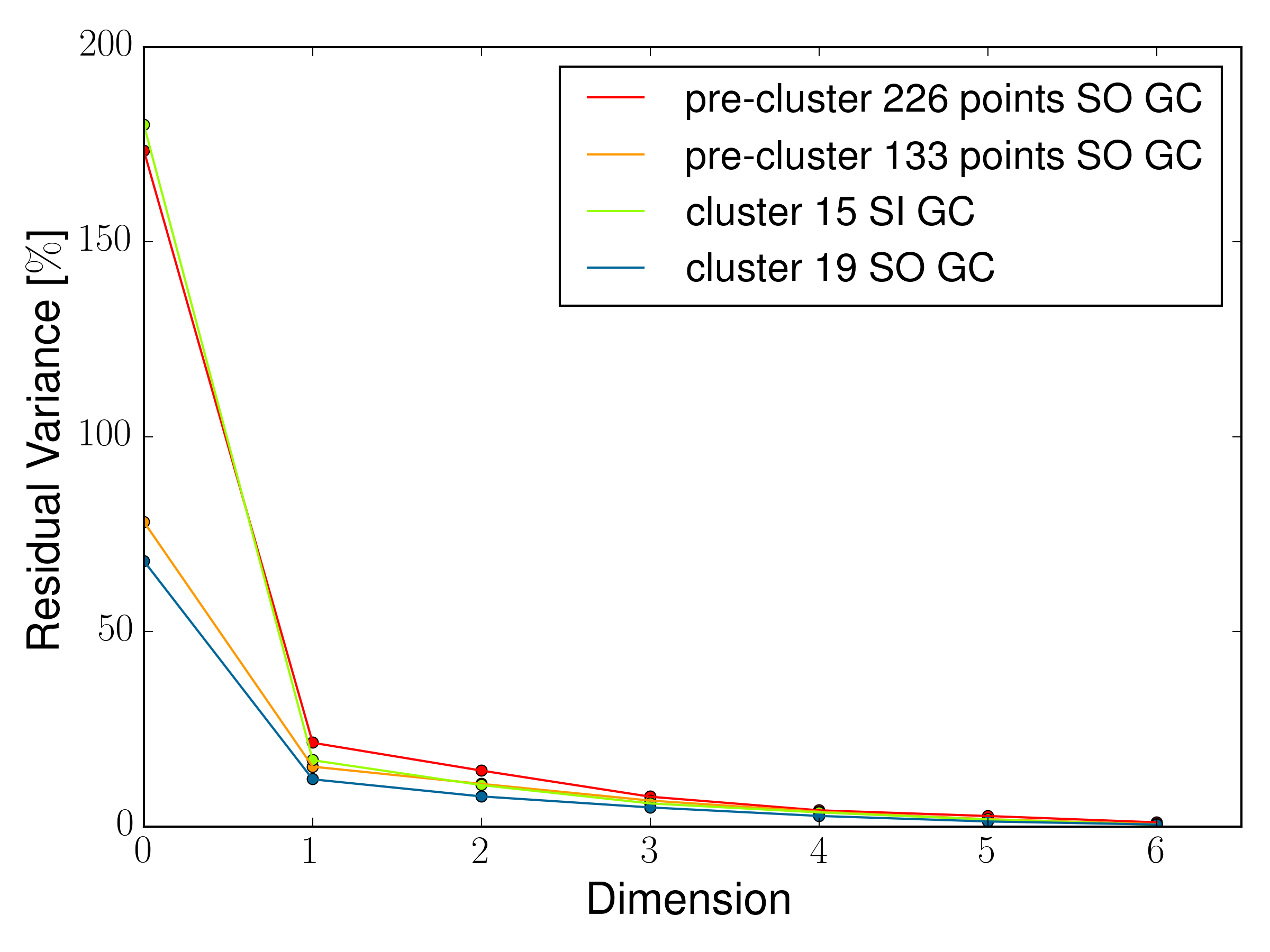}}
   \subcaptionbox{Tangent space PCA variances\label{clusters_screeb}}[0.45\textwidth]{\includegraphics[width=0.45\textwidth]{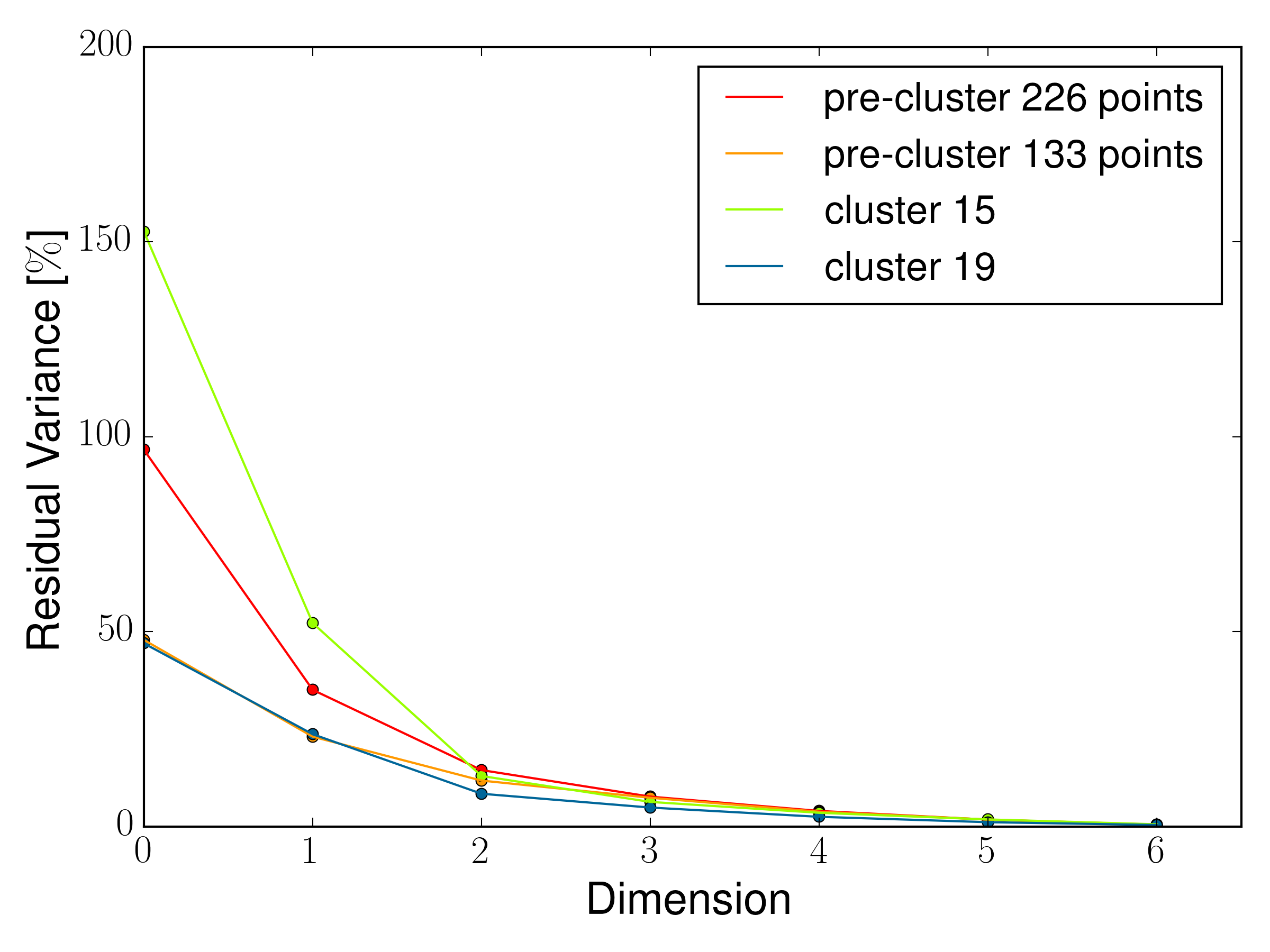}}
   \caption{\it Residual variance plots of pre-clusters: T-PCA (\ref{clusters_screea}) versus tangent space PCA (\ref{clusters_screeb}). These plots include only pre-clusters where the results differ markedly.}
   \label{clusters_scree}
\end{figure}

\paragraph{T-PCA method:} By pre-clustering we find $15$ clusters, which will in the following be called \textit{pre-clusters} and are listed in the Appendix. About $10\%$ of the data are characterized as outliers. In Figure \ref{clusters_screea} we give an illustration of the residual variances for pre-clusters 5, 8, 11 and 12 that shows the low residual variance of the one-dimensional T-PCA representation. For almost all pre-clusters, the one-dimensional T-PCA representation has less than $20\%$ residual data variance in relation to total data variance. The percentage at dimension zero for each pre-cluster gives its total variance relative to the large RNA data set's total variance as detailed in Section \ref{sec:variances}, in order to make non-Euclidean variances comparable with one-another.

Due to low residual variances of the one-dimensional projections of pre-clusters, we can use post-mode hunting and meaningfully interpret the found modes as clusters. This yields $22$ \textit{final clusters} with overall decreased variance and dimensionality; the smallest final cluster contains 28 points. Of the pre-clusters used in Figure \ref{clusters_screea}, pre-cluster 5 is decomposed into final clusters 11, 15 and 17 and pre-cluster 8 is decomposed into final clusters 14 and 18 while pre-clusters 11 and 12 remain unchanged as final clusters 16 and 19, respectively. Several of the final clusters, especially final clusters of low density, have apparently not been described before. Final clusters are also listed in the Appendix.

\begin{figure}[ht!]
  \centering
  \subcaptionbox{\it Pre-cluster 5 with 226 points\label{clusters_spheres-largea}}[0.45\textwidth]{\includegraphics[width=0.45\textwidth, clip=true, trim=0.5cm 0.2cm 0.5cm 0.2cm]{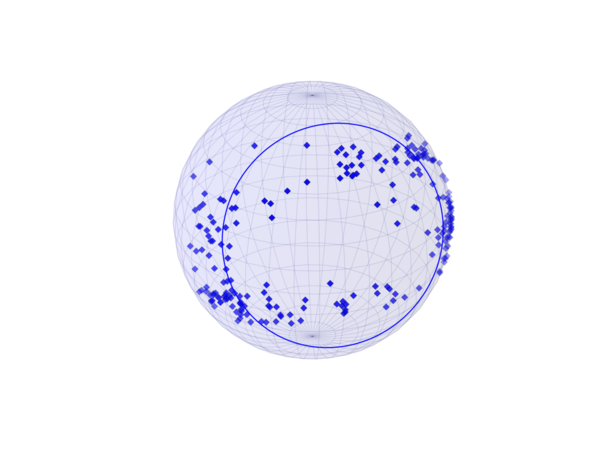}}
  \hspace*{0.05\textwidth}
  \subcaptionbox{\it Pre-cluster 12 with 52 points\label{clusters_spheres-largeb}}[0.45\textwidth]{\includegraphics[width=0.45\textwidth, clip=true, trim=0.5cm 0.2cm 0.5cm 0.2cm]{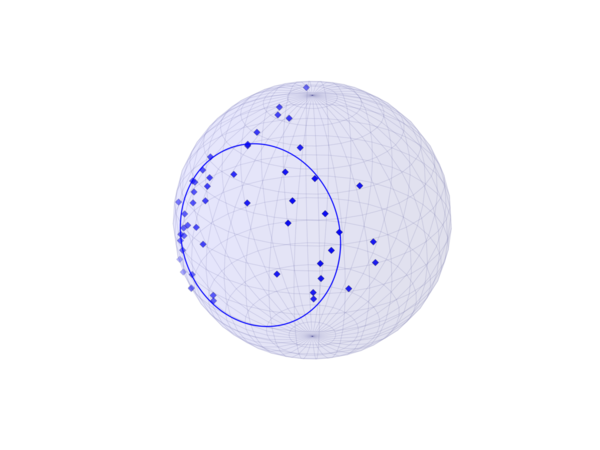}}
  \caption{\it Two-dimensional approximations of pre-clusters 5 and 12 from the large RNA data set for which T-PCA without mode hunting yields much better one-dimensional representations than tangent space PCA. The circles on the spheres illustrate the best fit circles found by T-PCA, along which mode hunting is performed.}
  \label{clusters_spheres-large}
\end{figure}

\paragraph{Comparison with tangent space PCA:} We find that for the pre-clusters 5, 8, 11 and 12, as used above for the T-PCA, tangent space PCA leaves more than $30\%$ residual data variance in the one-dimensional representation (see Figure \ref{clusters_screeb}), whereas the one-dimensional T-PCA representation has less than $20\%$ residual data variance. The root of the success of T-PCA over tangent space PCA can be seen by investigating those pre-clusters which are much better approximated by T-PCA. All of these pre-clusters have distinct non-linear shapes, so they are badly fitted by linear subspaces. Figure \ref{clusters_spheres-large} displays two-dimensional projections of pre-clusters 5 and 12, whose residual variances are displayed in Figure \ref{clusters_scree}. In particular the non-linear shapes are preserved in the two-dimensional representation and one can clearly see that they are one-dimensionally much better fit by small circles which is impossible (let alone the two-dimensional spherical representation) via a tangent space approach.

\subsubsection{Locating a New Low Density Cluster}\label{sec:data-clustering}

We now illustrate the power of our method by example of using three final clusters from the large RNA data set. These clusters are numbers $1$, $2$ and $7$ out of the $22$ final clusters listed in the Appendix (Tables 1 and 2); these contain a total of $5625$ out of $8301$ data points. These three clusters have been selected because they strongly overlap and are inseparable in the 2D pseudo-torsion representation. To large parts, the high density clusters $1$ and $2$ have been described by \cite{Richardson2008}. Cluster $7$ has low density and could only be found by T-PCA.

\begin{figure}[ht!]
   \centering
   \subcaptionbox{$\eta$--$\theta$ plot\label{clusters_overlapa}}[0.45\textwidth]{\includegraphics[width=0.45\textwidth, clip=true, trim=2cm 0 2cm 0]{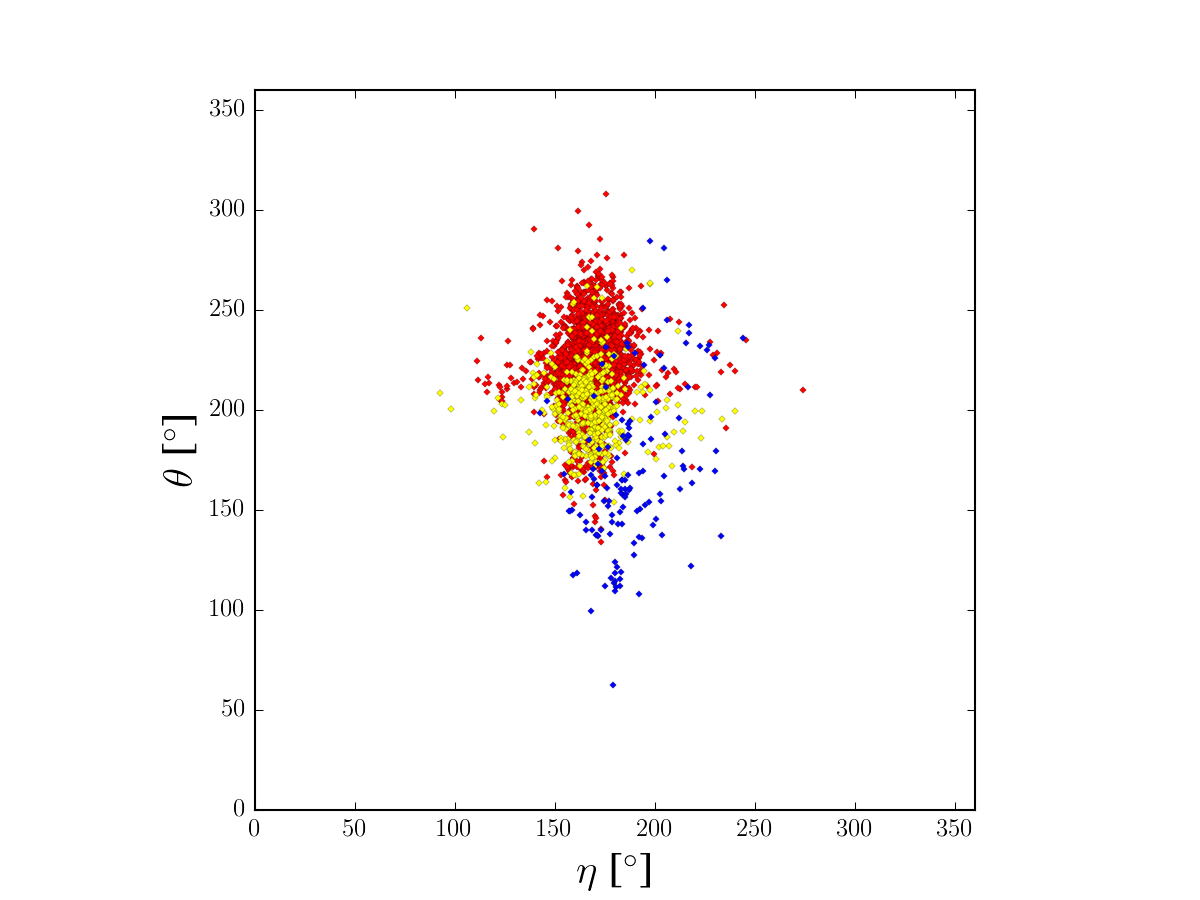}}
   \hspace*{0.05\textwidth}
   \subcaptionbox{$2d$ T-PCA subsphere plot\label{clusters_overlapb}}[0.45\textwidth]{\includegraphics[width=0.4\textwidth, clip=true, trim=0.55cm 0.25cm 0.5cm 0.25cm]{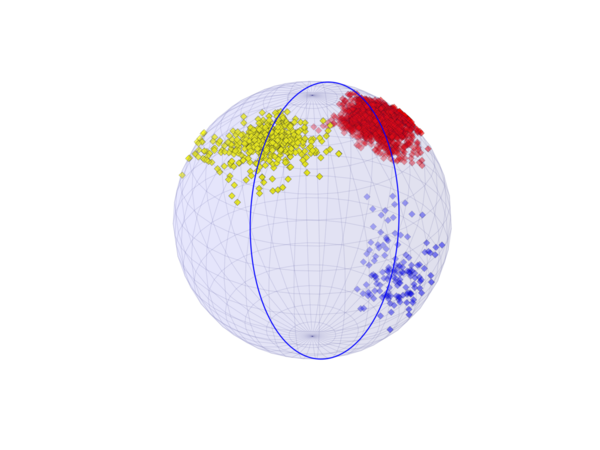}}
   \caption{\it Clusters $1$ in red, $2$ in yellow and $7$ in blue. They overlap in the $\eta$--$\theta$ plot (\ref{clusters_overlapa}) but can easily be separated in the 7D torus of $\alpha$-$\beta$-$\gamma$-$\delta$-$\epsilon$-$\zeta$-$\chi$ by T-PCA, as illustrated by the two-dimensional projection displayed in \ref{clusters_overlapb}.}
   \label{clusters_overlap}
\end{figure}

\begin{table}[!ht]
  \caption{\textit{Properties of clusters $1$ (red), $2$ (yellow) and $7$ (blue) of Figure \ref{clusters_overlap}. The ``\# Bonds'' section of Table \ref{table_cluster_statsa} gives the relative number of bases bound to $1$, $2$ or $3$ other bases. Base pair types in Table \ref{table_cluster_statsb} are denoted by five letters representing in this order: 1.~bond region of the base, 2.~bond region of its partner, 3.~cis/trans bond geometry, 4.~type of the base, 5.~type of its partner. The bond regions are denoted as follows: W for Watson-Crick, H for Hoogsteen, S for Sugar. For example, ``\textbf{SHtGA}'' means that the sugar edge (\textbf{S}) of a cluster residue is bound to the Hoogsteen edge (\textbf{H}) of another residue with trans (\textbf{t}) alignment, where the cluster residue is a Guanine (\textbf{G}) base and its bond partner is an Adenine base (\textbf{A}). The information in the ``\# Bonds'' section of Table \ref{table_cluster_statsa} and all of Table \ref{table_cluster_statsb} was extracted using the proprietary RNAview software, see \cite{Yang2003}.}}
  \subcaptionbox{General information\label{table_cluster_statsa}}[0.49\textwidth]{ 
    \begin{tabular}{|l|r|r|r|}
      \hline
      \textbf{Cluster \#} & $1$ & $2$ & $7$ \\ \vspace*{0.5\baselineskip}
      \textbf{\# Points} & $4921$ & $477$ & $137$ \\ \hline \vspace*{0.5\baselineskip}
      C3'-endo & $100 \%$ & $100 \%$ & $3.62 \%$ \\ \hline \vspace*{0.5\baselineskip}
      \textbf{Bases} &&& \\ \hline \vspace*{0.5\baselineskip}
      A & $19.89 \%$ & $16.35 \%$ & $25.36 \%$ \\ \vspace*{0.5\baselineskip}
      C & $31.64 \%$ & $24.74 \%$ & $11.59 \%$ \\ \vspace*{0.5\baselineskip}
      G & $32.55 \%$ & $46.75 \%$ & $42.75 \%$ \\ \vspace*{0.5\baselineskip}
      U & $15.91 \%$ & $12.16 \%$ & $20.29 \%$ \\ \hline \vspace*{0.5\baselineskip}
      \textbf{\# Pairs} &&& \\ \hline \vspace*{0.5\baselineskip}
      $1$ & $71.88 \%$ & $67.92 \%$ & $44.20 \%$ \\ \vspace*{0.5\baselineskip}
      $2$ & $17.84 \%$ & $18.66 \%$ & $30.43 \%$ \\ \vspace*{0.5\baselineskip}
      $3$ & $3.13 \%$ & $5.03 \%$ & $4.35 \%$ \\ \hline
    \end{tabular}
  }
  \subcaptionbox{Bond information\label{table_cluster_statsb}}[0.49\textwidth]{ 
    \begin{tabular}{|l|r|r|r|}
      \hline
      \textbf{Cluster \#} & $1$ & $2$ & $7$ \\ \vspace*{0.5\baselineskip}
      \textbf{\# Points} & $4921$ & $477$ & $137$ \\ \hline \vspace*{0.5\baselineskip}
      \textbf{Bonds} &&& \\ \hline \vspace*{0.5\baselineskip}
      WWcCG & $26.91 \%$ & $20.13 \%$ & $ 5.07 \%$ \\ \vspace*{0.5\baselineskip}
      WWcGC & $23.15 \%$ & $38.16 \%$ & $ 7.25 \%$ \\ \vspace*{0.5\baselineskip}
      WWcAU & $ 7.88 \%$ & $ 4.61 \%$ & $ 1.45 \%$ \\ \vspace*{0.5\baselineskip}
      WWcUA & $ 8.01 \%$ & $ 4.40 \%$ & $ 0.72 \%$ \\ \vspace*{0.5\baselineskip}
      SHtGA & $ 1.95 \%$ & $ 1.68 \%$ & $16.67 \%$ \\ \vspace*{0.5\baselineskip}
      SHcAA & $ 0.08 \%$ & $ 0.00 \%$ & $ 5.07 \%$ \\ \vspace*{0.5\baselineskip}
      HWtAU & $ 0.33 \%$ & $ 1.68 \%$ & $ 4.35 \%$ \\ \vspace*{0.5\baselineskip}
      WWcGU & $ 2.84 \%$ & $ 4.19 \%$ & $ 0.72 \%$ \\ \vspace*{0.5\baselineskip}
      none  & $ 6.77 \%$ & $ 8.18 \%$ & $21.01 \%$ \\ \hline
    \end{tabular}
  }
  \label{table_cluster_stats}
\end{table}

The molecular properties assembled in Tables \ref{table_cluster_stats} show that cluster $7$ (blue in Figure \ref{clusters_overlap}) is very different from the other two clusters (cluster $1$ in red and cluster $2$ in yellow in Figure \ref{clusters_overlap}). One of the most striking differences, see Table \ref{table_cluster_statsa}, is that clusters $1$ and $2$ consist only of residues with C3'-endo sugar pucker, while cluster $7$ features mostly the alternate C2'-endo sugar pucker. Furthermore, residues in cluster $7$ have mostly irregular or no bonds (all bonds not of base pair type ``WWc$\cdots$'' i.e. below the first four in Table \ref{table_cluster_statsb} are irregular), and display an abundance of bases connected to more than one other base. Clusters $1$ and $2$ are very similar in terms of bond patterns although these differ most strikingly by the excess of Guanine residues in cluster $2$ of $47 \%$ versus $33 \%$, see Table \ref{table_cluster_statsa}, which is mirrored by an excess of corresponding bonds.

\begin{figure}[ht!]
   \centering
   \begin{minipage}{0.4\textwidth}
   \includegraphics[width=1\textwidth,clip=true, trim = 10 0 6 6]{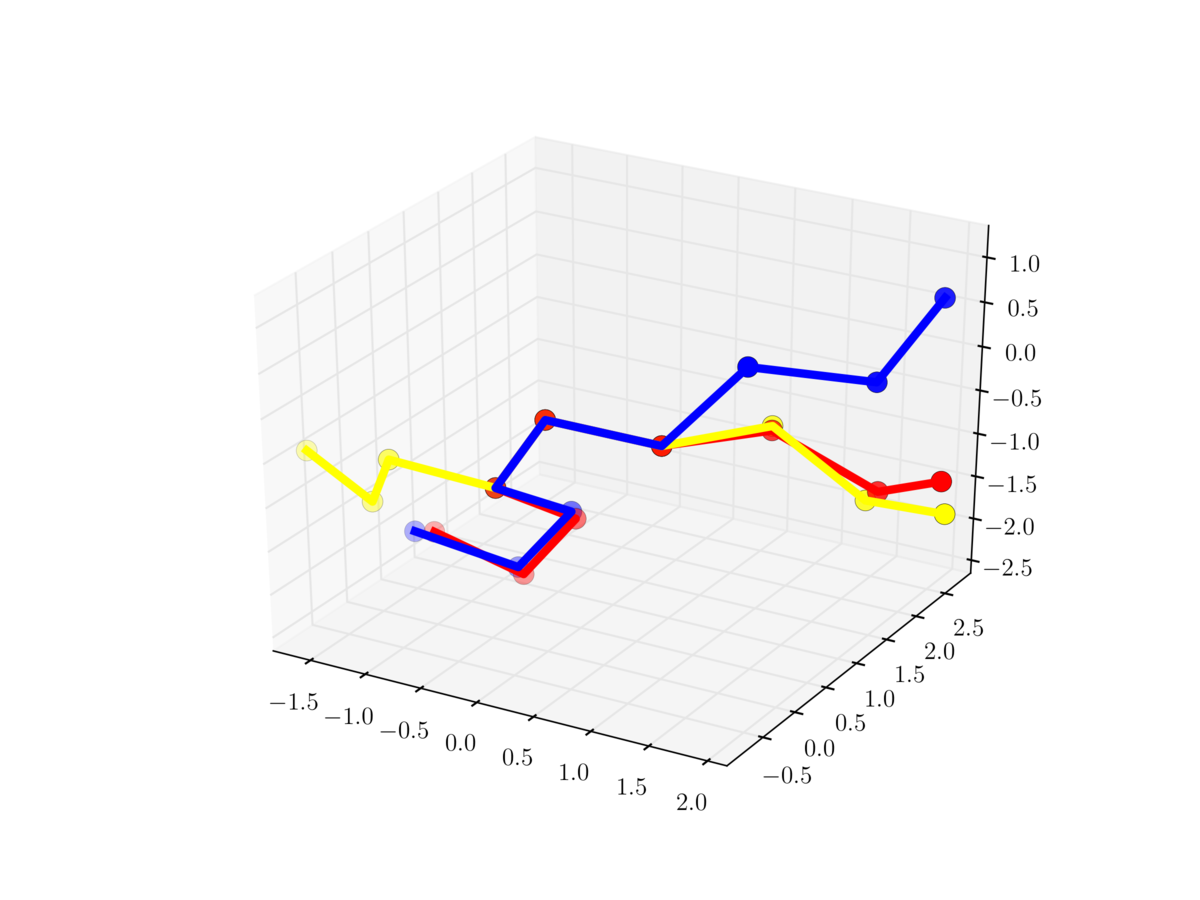}
   \end{minipage}
   \begin{minipage}{0.03\textwidth}\hfill
   \end{minipage}
   \begin{minipage}{0.55\textwidth}
   \caption{\it Atomic geometries of the backbone pieces corresponding to the DT-PNS nested means for cluster $1$ (red), $2$ (yellow) and $7$ (blue) with same colors as in Figure \ref{clusters_overlap}. This display is analogous to Figure \ref{rna_backbone_3d}. The first three and last three atoms represent the phosphate groups and the chains are aligned along the three carbon atoms from the sugar to visualize similarities between the structures.  (For clear vizualization we have used constant bond length.)\label{residues_geometry}}
   \end{minipage}

\end{figure}

The only dihedral angles for which nested mean values differ significantly among cluster $1$ and $2$ are $\alpha$ and $\gamma$ while the nested mean of cluster $7$ deviates from that of cluster $1$ in the angles $\delta$ and $\zeta$. For cluster $1$, $\alpha \approx 297^\circ$, $\gamma \approx 53^\circ$, $\delta \approx 81^\circ$ and $\zeta \approx 293^\circ$, while for cluster $2$, $\alpha \approx 154^\circ$ and $\gamma \approx 176^\circ$ and for cluster $7$, $\delta \approx 147^\circ$ and $\zeta \approx 149^\circ$. Thus, residues from both clusters $2$ and $7$ can be regarded as kinked versions of the residues of cluster $1$ as illustrated in Figure \ref{residues_geometry}. Our findings concerning $\delta$ seem well in accordance with previous results, e.g. \cite{Richardson2008} note that the C3'-endo pucker corresponds to a mean $\delta$ between $78^\circ$ and $90^\circ$ while the C2'-endo pucker leads to mean $\delta$ between $140^\circ$ and $152^\circ$. The sugar puckers explain that cluster $2$ is larger than cluster $7$, see Table \ref{table_cluster_statsa}.

\subsection{Helical Structures}

\begin{figure}[ht!]
   \centering
   \subcaptionbox{Helical conformations\label{helicesa}}[0.45\textwidth]{\includegraphics[width=0.49\textwidth,clip=true, trim=5 2 5 5]{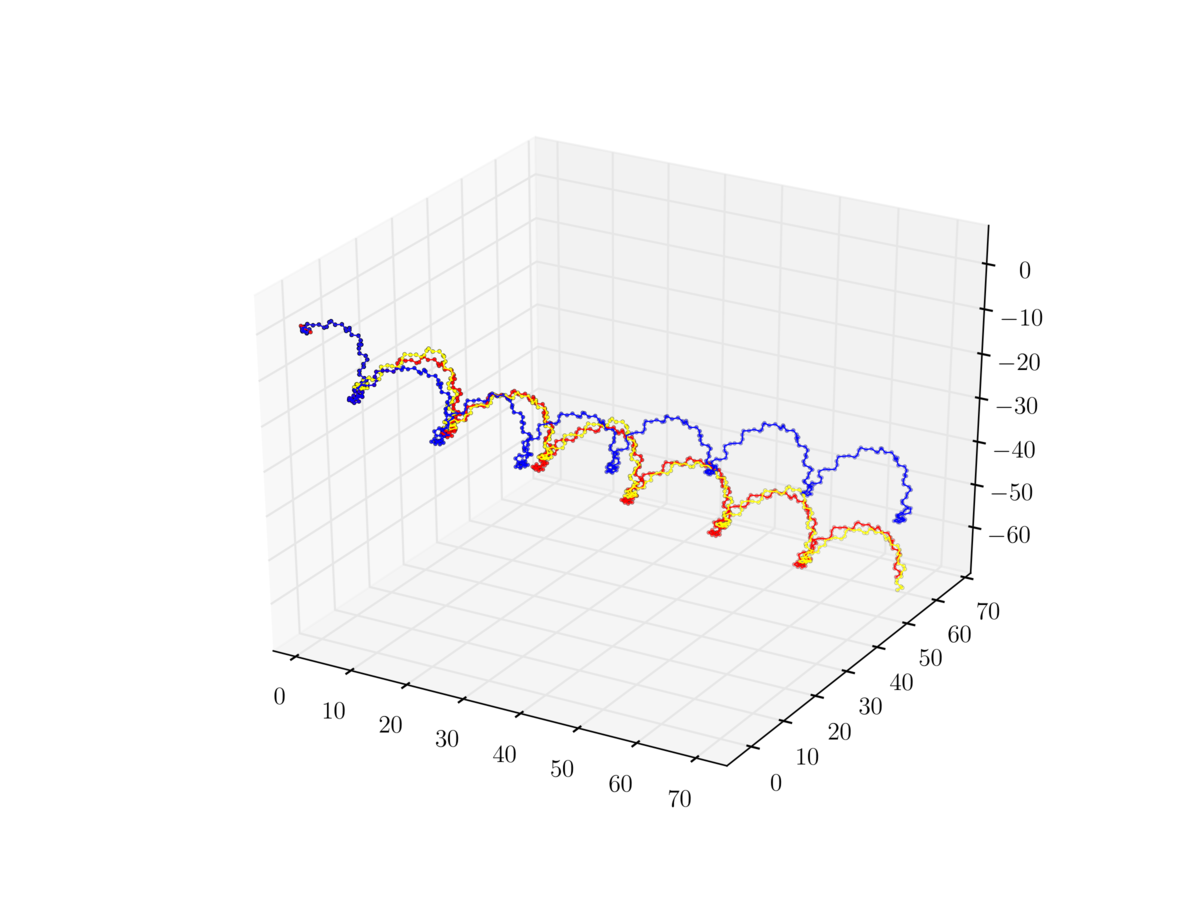}}
   \subcaptionbox{Irregular strands\label{helicesb}}[0.45\textwidth]{\includegraphics[width=0.49\textwidth,clip=true, trim=5 2 5 5]{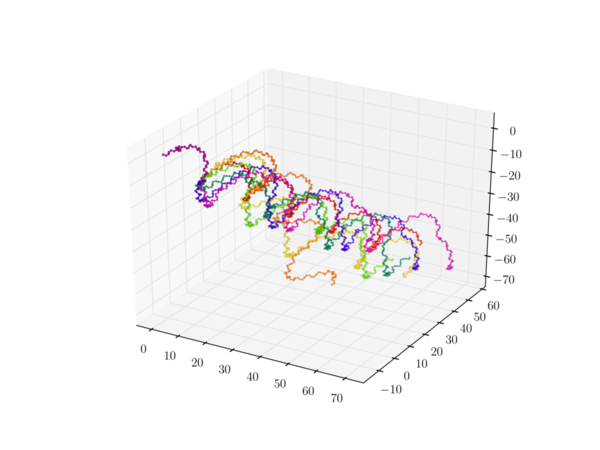}}
   \caption{\it Strands of typical residues of clusters $1$, $2$ and $7$. (Coloring scheme here is independent of Figures \ref{clusters_overlap} and \ref{residues_geometry}.) \ref{helicesa}: a sequence solely of cluster 1 residues (red); a sequence of mostly cluster 1 residues, every sixth residue being a cluster 2 residue (yellow); a sequence of mostly cluster 1 residues, every 11th residue being a cluster 2 residue (blue). \ref{helicesb}: Conformations of cluster 1 residue sequences interspersed with cluster 7 residues.}
   \label{helices}
\end{figure}

We continue with the three clusters of Section \ref{sec:data-clustering}. To describe the typical residue geometry of a cluster, we use its nested mean, as defined in Section \ref{subsec:dt-pca}, in the following. A backbone consisting only of typical residues from cluster $1$ takes a helical shape with approximately $11$ residues forming one turn, which is typical for the A-helix conformation frequently found in RNA structures, see Figure \ref{helicesa} and \cite{Duarte1998,Wadley2007,Richardson2008}. It is not possible to form such structures consisting of typical residues either alone from cluster $2$ or alone from cluster $7$ as these result in a tightly wound helix, which is incompatible with bases attached to the residues. Interspersing a backbone of typical residues from cluster $1$, however, with typical residues of cluster $2$ leads to a variety of bent or loose helical shapes, see Figure \ref{helicesa}. This is consistent with the often irregular shapes of RNA molecules, see e.g. \cite{Wadley2007}. Typical residues from cluster $7$ when interspersed in a backbone consisting mostly of typical residues from cluster $1$ result in very irregular strands, some of which are depicted in Figure \ref{helicesb}. These seem not frequently observed as larger secondary structure, however. This suggests that such mixed conformations are rather unusual in long strands.

Indeed, this is confirmed by our investigation of the RNA chains of the large RNA data set we use, where we identify contiguous sequences of residues from the three clusters investigated here. The sequences containing residues from cluster $7$ are much shorter on average than those containing residues from cluster $2$. Furthermore, the residues from cluster $7$ are much less likely to be located in the middle of a sequence than those from cluster $2$, indicating that they are not usually part of rather regular helical regions.

In summary, clusters $1$ and $2$ can be clearly associated with helical structure elements, while cluster $7$, which is clearly a distinct cluster in the data, does not correspond alone to a typical structural element. 

\section{Discussion}

We have provided a novel framework for torus PCA to perform PCA-like dimension reduction for angular data. Previous attempts have not been satisfactory, because, on the one hand, the geometry featuring dense geodesics lead to severe restrictions for geodesic approaches while, on the other hand, Euclidean approximations disregard periodicity. We have used an adaptive deformation to a benign geometry, whilst at the same time preserving periodicity. For T-PCA to be fully effective it needs pre- and post-clustering. In application to dihedral angles of RNA structures we validated our method using a small classical benchmark data set. On a large classical data set, we go well beyond results achieved by analysis of 2-dimensional pseudo-torsion angles or recent 7-d clustering methods. Also we provide moderately sized clusters, half of them of size between $59$ and $139$ points. In fact, we have identified several clusters of low density which have not been located before. Some clusters have been examined in relation to helical conformations. Our method is widely applicable and can be used for geometrical analysis of biomolecular strands such as proteins, DNA and others.    

\section*{Acknowledgements}

We are grateful to Thomas Hamelryck, John Kent and J.S. Marron for helpful discussions. We thank Jes Frellsen for his comments on a draft of this article and for pointing to the RNAview program for calculation of RNA bonds. Further, we wish to thank Karen Sargsyan for providing details on GeoPCA and on the RNA data used by them in their paper.

\appendix
\section{Supplementary material}

\subsection{Additional Illustrations}

\begin{figure}[ht!]
  \centering
  \includegraphics[width=0.95\textwidth]{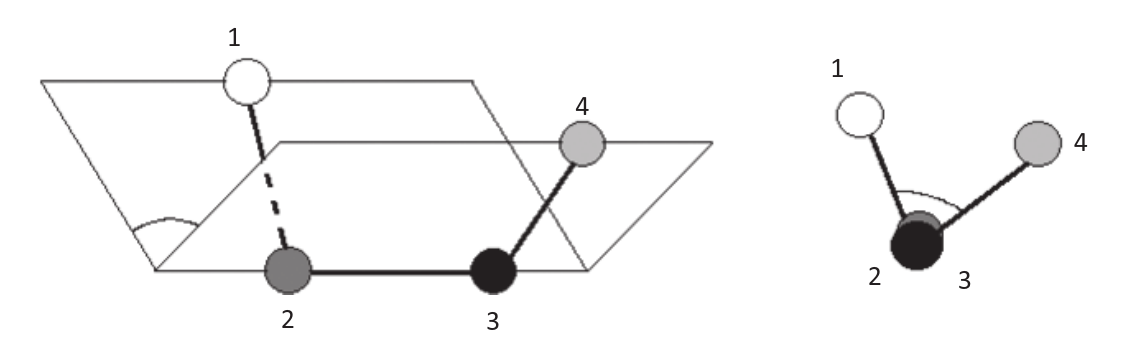}
  \caption{\it Illustration of a dihedral angle defined by four atoms or three bonds, it is the opening angle between to pages of a book. (Reproduced from \cite{Mardia2013}.)}
  \label{dihedral_angles}
\end{figure}

\FloatBarrier

\subsection{Polar Coordinates for Higher Dimensions}

Assuming the embedding $\mathbb{S}^D=\{x\in \mathbb R^{D+1}:\|x\|=1\}$, the coordinates of the embedding space $x_k$ are related to angular coordinates $\phi_k$ as follows
\begin{align*}
  x_1 &= \cos \phi_1\\
  \forall 2 \le k \le D \, : \, x_k &= \left( \prod_{j=1}^{k-1} \sin \alpha_j \right) \cos \phi_k\\
  x_{D+1} &= \left( \prod_{j=1}^{D} \sin \phi_j \right) .
\end{align*}

\subsection{Flow Chart of the T-PCA Algorithm}

\begin{figure}[ht!]
   \centering
   \includegraphics[width=0.95\textwidth]{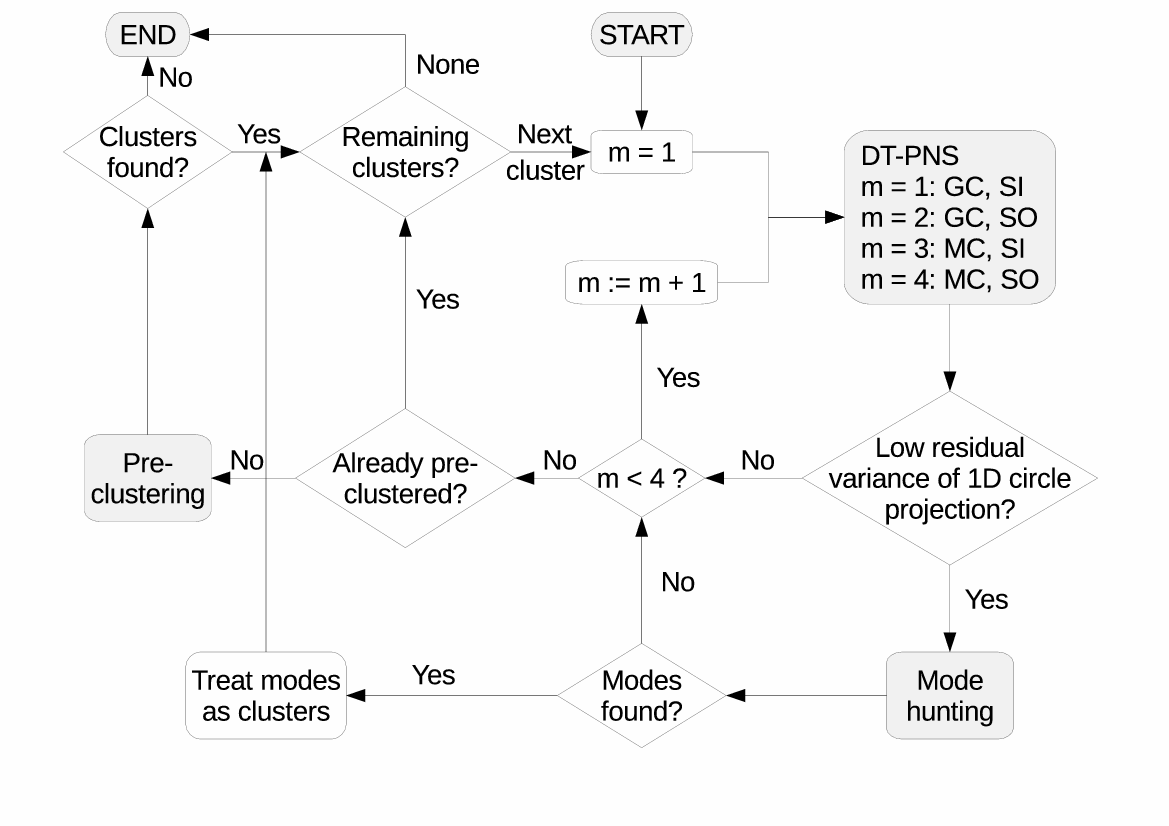}
   \caption{\it The torus-PCA algorithm including DT-PNS, pre-clustering and mode hunting.}
   \label{flow_chart}
\end{figure}

\FloatBarrier

\subsection{Abbreviations and Technical Terms}

We give a brief overview over abbreviations and technical terms used throughout this paper:\\
\textbf{T-PCA}:~Torus Principal Component Analysis. The dimension reduction method via geometrical deformation presented in the present article.\\
 \textbf{DT-PNS}:~Deformed Torus Principal Nested Spheres, an alteration of PNS by \cite{Jung2012}. A backwards method for dimension reduction on spheres. At each step, one finds the small subsphere with codimension $1$ which best fits the data.\\
\textbf{MC}:~Mean~Centered, \textbf{GC}:~Gap~Centered, \textbf{SI}:~Spread~Inside, \textbf{SO}:~Spread~Outside,\\ \textbf{H}:~Halved~angles, \textbf{U}:~Unscaled~angles, see Subsection \ref{sec:sausage_surg}.\\
\textbf{Codimension}: The codimension of $k$-dimensional subspace of a $d$-dimensional space is defined as $d-k$.\\
\textbf{Residual}: Statistically unexplained data variation, see Subsection \ref{sec:variances}.\\
\textbf{Residue}: RNA molecule segment corresponding to a single nucleic base, see Subsection \ref{sec:appl_rna}.
\textbf{Nested Mean}: The ultimate point $\mu$ of the sequence of small subspheres $\mathbb{S}^D \supset S^{D-1} \supset \dots \supset S^2 \supset S^1 \supset \{ \mu\}$ found by PNS, see subsection \ref{subsec:dt-pca}.

\subsection{Topological Details}\label{sec:scaling-and-topology}

\begin{figure}[ht!]
  \centering
  \includegraphics[width=0.3\textwidth, clip=true, trim=1cm 0.8cm 0.8cm 0.6cm]{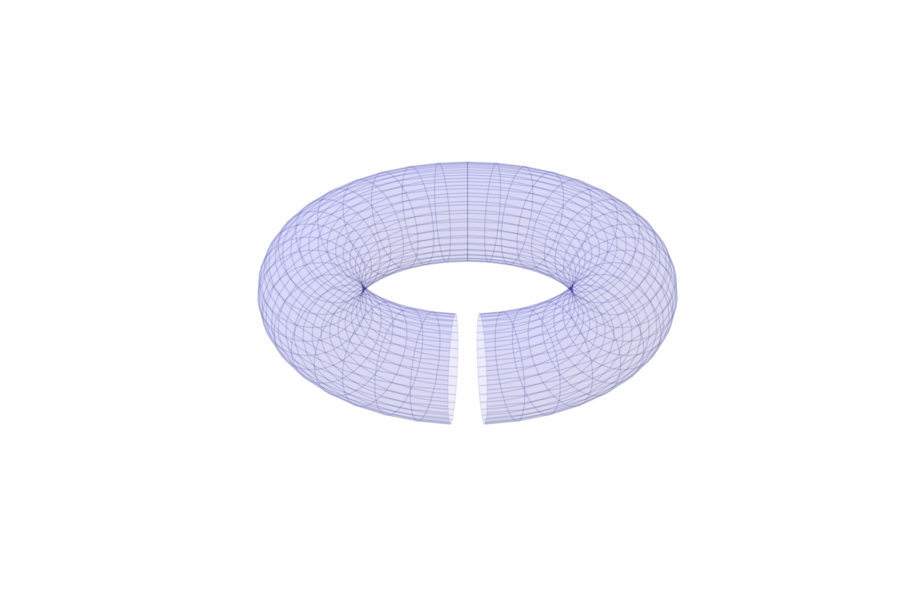}
  \hspace*{0.01\textwidth}
  \includegraphics[width=0.3\textwidth, clip=true, trim=1cm 0.8cm 0.8cm 0.6cm]{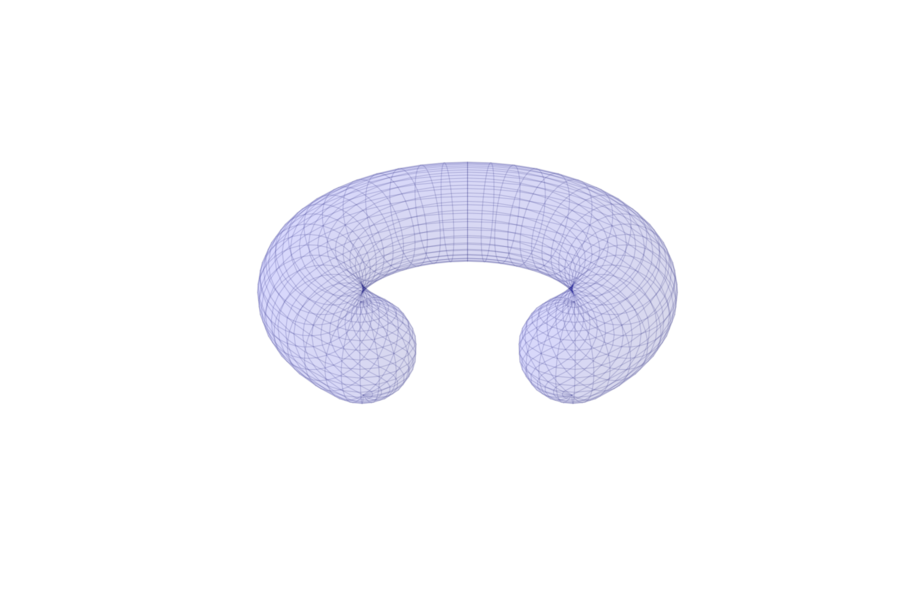}
  \hspace*{0.01\textwidth}
  \includegraphics[width=0.3\textwidth, clip=true, trim=1cm 0.8cm 0.8cm 0.6cm]{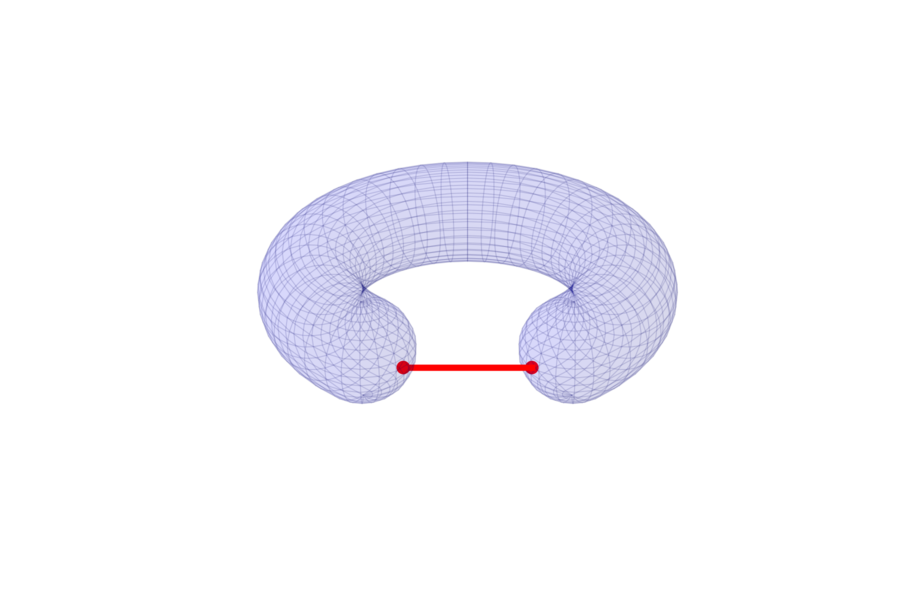}
  \caption{\it Gluing in the case of halved angles for $T^2$: the introduced curvature due to embedding $T^2$ in $\mathbb{R}^3$, for illustration's sake, should not be mistaken for the torus deformation of Section \ref{sec:sausage_surg}}.
  \label{sausage-deformation}
\end{figure}

\begin{figure}[ht!]
  \centering
  \includegraphics[width=0.3\textwidth, clip=true, trim=2cm 0cm 2cm 0cm]{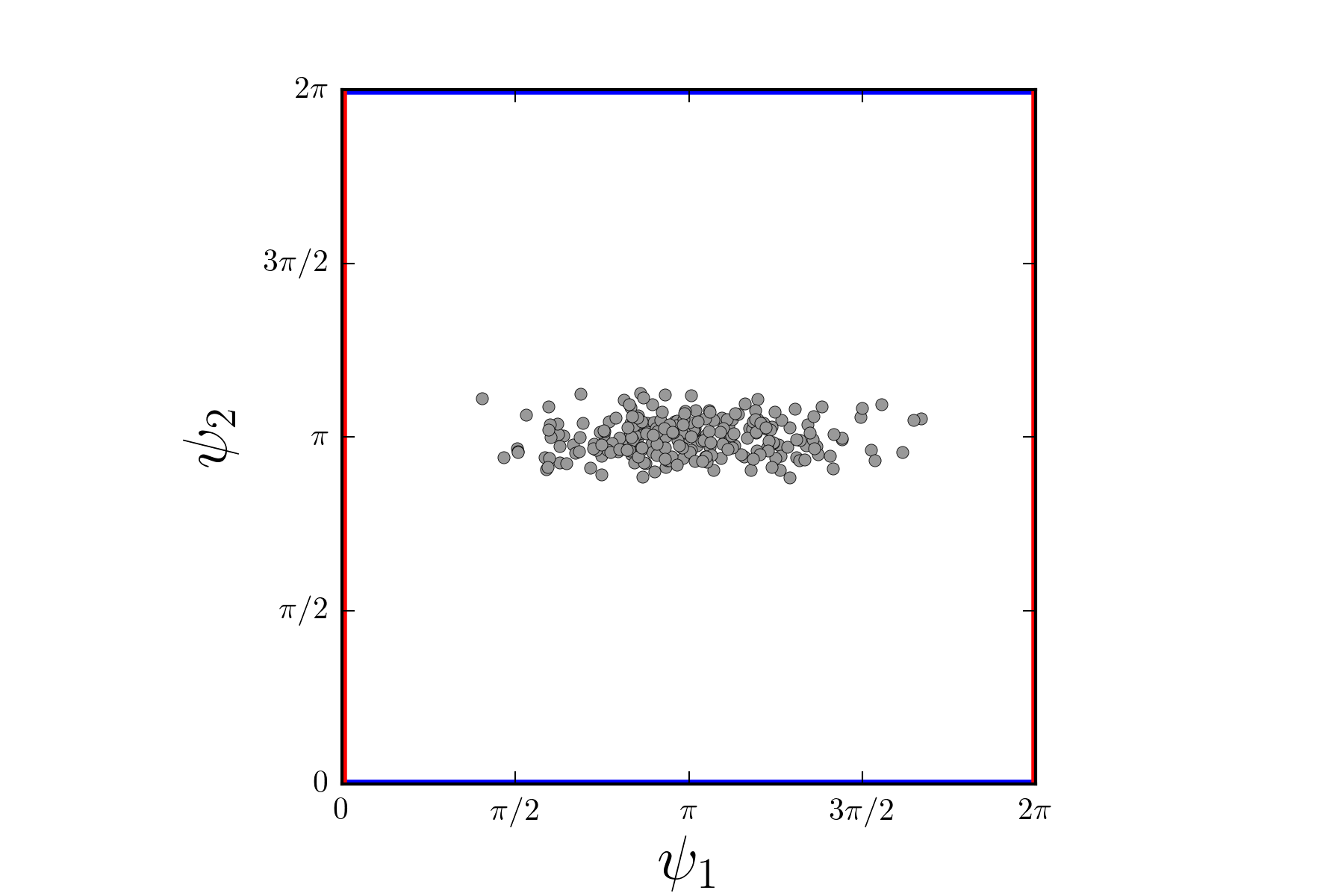}
  \hspace*{0.01\textwidth}
  \includegraphics[width=0.3\textwidth, clip=true, trim=2cm 0cm 2cm 0cm]{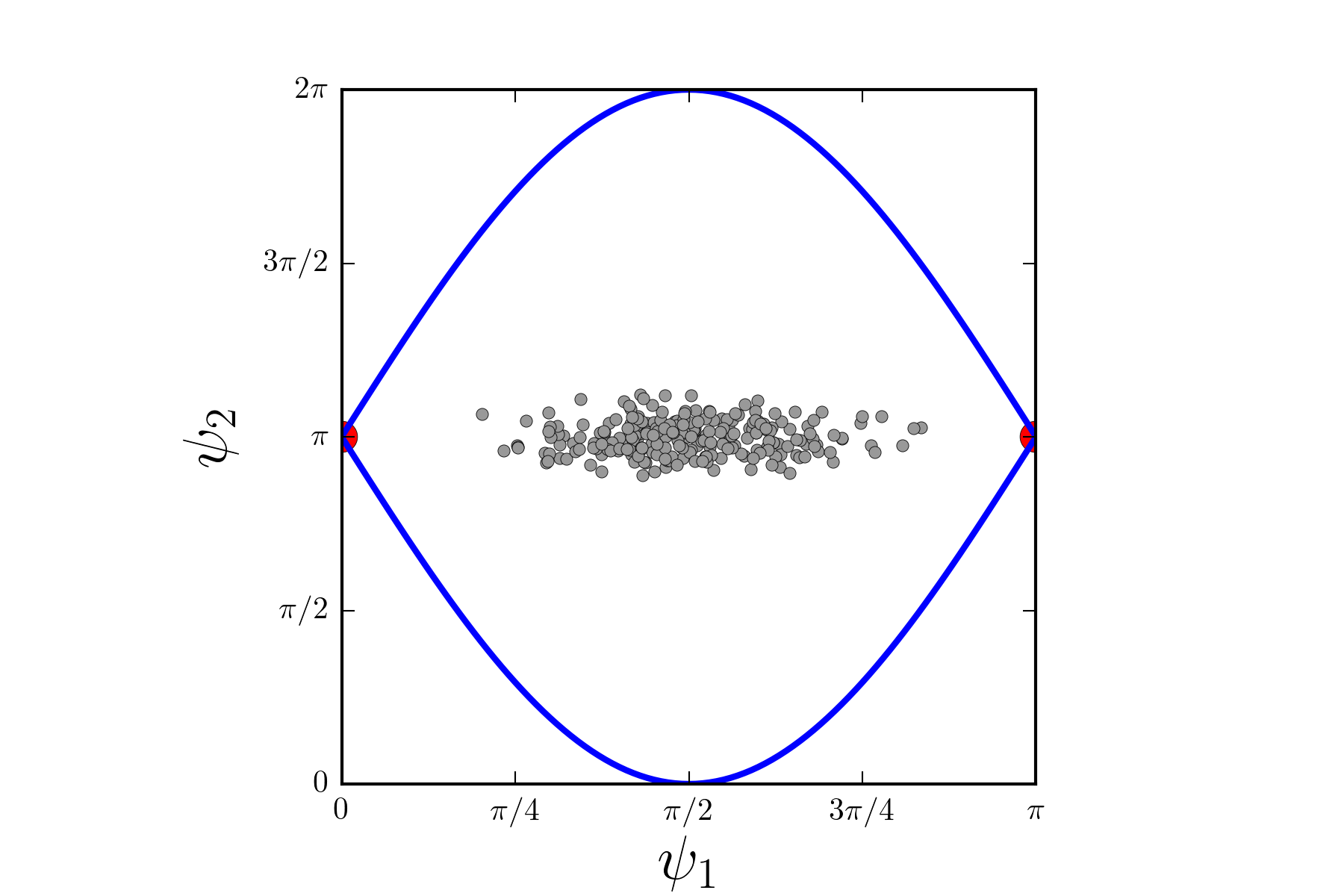}
  \hspace*{0.01\textwidth}
  \includegraphics[width=0.3\textwidth, clip=true, trim=1cm 0.5cm 1cm 0.5cm]{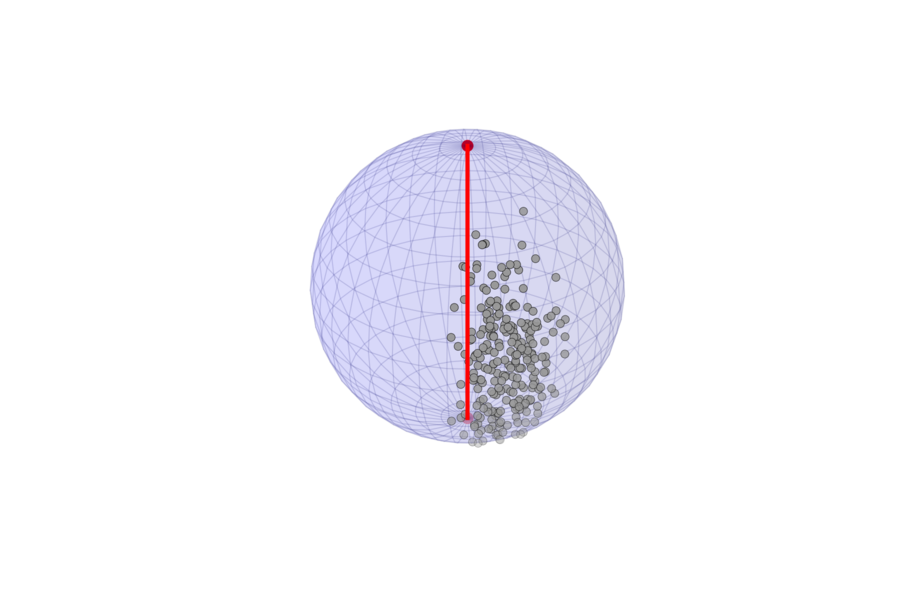}\\
  \hspace*{0.32\textwidth}
  \includegraphics[width=0.3\textwidth, clip=true, trim=2cm 0cm 2cm 0cm]{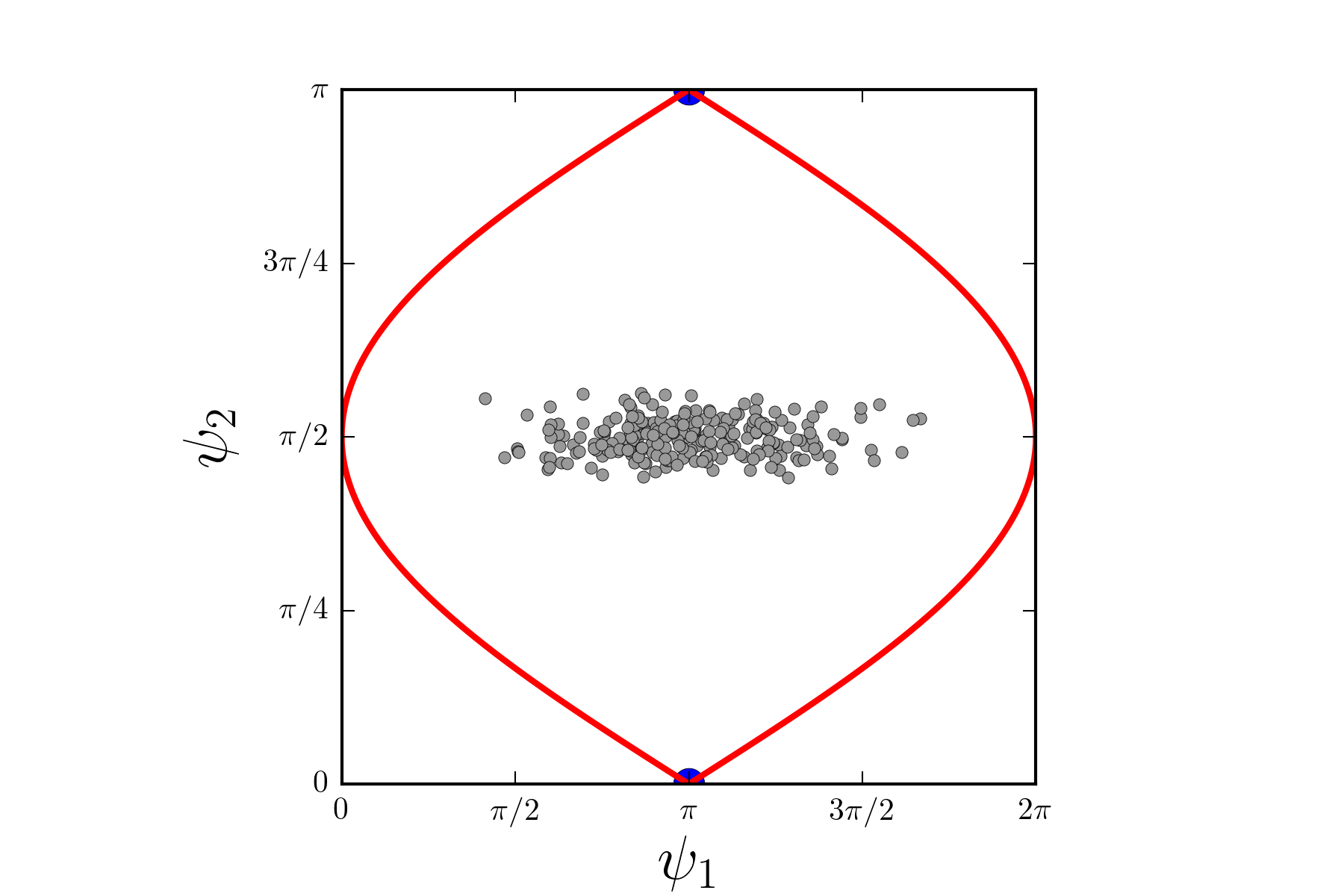}
  \hspace*{0.01\textwidth}
  \includegraphics[width=0.3\textwidth, clip=true, trim=1cm 0.5cm 1cm 0.5cm]{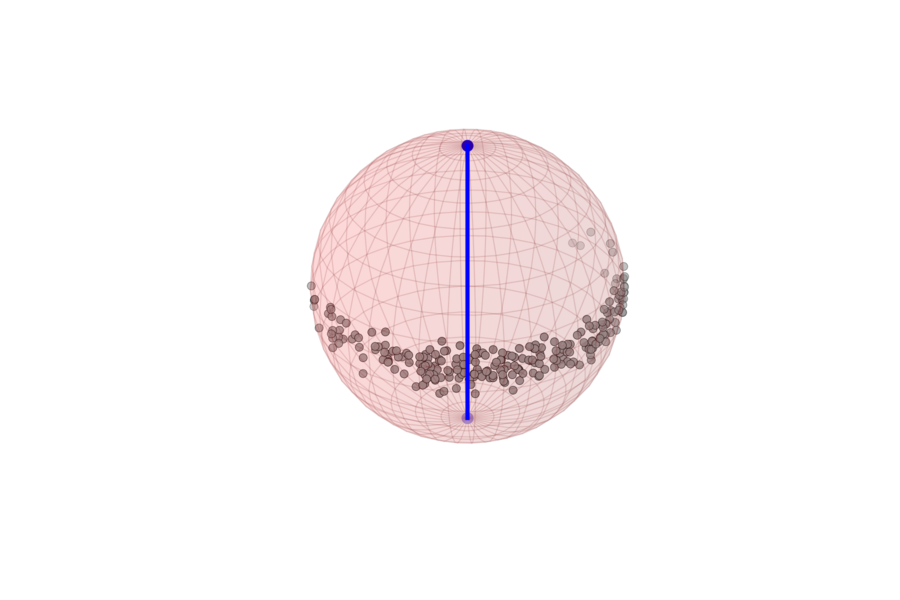}
  \caption{\it Two possibilities for gluing in the case of halved angles for $T^2$. Top row: data on a torus in flat representation and the effect of halving $\psi_1$; due to the torus' periodicity, top and bottom blue arcs are identified; due to collapsing of the identified red lines to points (the singularity set), north and south pole of the sphere are identified (shown by the red line). Bottom row:  the effect of halving $\psi_2$ with the roles of red and blue reversed.}
  \label{torus-halving-deformation}
\end{figure}

Due to periodicity on the torus, $\psi_k=0$ is identified with $\psi_k=2\pi$ for all $k=1,\ldots,D$. In contrast, for all angles $\phi_k=0$ denotes spherical locations different from $\phi_k=\pi$. In case of halving (H), except for the innermost angle ($1\leq k< D$), for an invariant representation respecting torus distance, however, it is necessary to identify these locations accordingly, which results in a self-gluing of $\mathbb{S}^D$ along specific codimension two great subspheres, see Figures \ref{sausage-deformation} and \ref{torus-halving-deformation}.  

In the rare case of angles concentrated to an interval of length $\pi$, using unscaled angles (U) this interval is mapped to $[0,\pi]$ without any distortion.

%

Here is an illustration for the gluing effects in the case of halving. 

\begin{Ex} For $D=3$, on $\mathbb{S}^3$ we have the squared line element
\begin{align*}
  ds^2 = d\phi_1^2 + \sin^2 \phi_1 \left( d\phi_2^2 + \sin^2 \phi_2 d\phi_3^2 \right)\,.
\end{align*}
where the angle ranges are $\phi_{1},\phi_{2} \in (0,\pi), \, \phi_3 \in [0,2\pi)$.
  When using halved angles, for $\phi_1$ and $\phi_2$ we have the identification $0 \equiv \pi$. For $\phi_1$ this is an identification of two points. For $\phi_2$ this is an identification of the points $(\phi_1, 0, 0)$ and $(\phi_1, \pi, 0)$ for all $\phi_1 \in (0, \pi)$, which means that pairs of points are glued together along half circles. The example $D=2$ is illustrated in Figures \ref{sausage-deformation} and \ref{torus-halving-deformation}.
\end{Ex}

\subsection{Improved Hypothesis Test}

Let $S^d$ be a fitted small subsphere, $2\leq d\leq D$. For ease of notation, assume that $S^d = \mathbb{S}^d$ and that $p \in \mathbb{S}^d$ is the center of the fitted small $S^{d-1}$. For simplicity, we restrict attention to probability distributions $q \mapsto g(q;p)$ which depend only on the angular distance $r = d(p,q)$ for any data point $q \in \mathbb{S}^d$. Let $\gamma$ be a curve along any great circle connecting $p$ with its antipodal and let $\gamma$ be parametrized by $r \in [0, \pi]$ such that $\forall r : \, d(p,\gamma(r)) = r$. Then, due to its symmetry, $g$ is fully characterized by the function
\begin{align*}
  h(r;p) := \text{vol}_{\mathbb{S}^{d-1}} \cdot g(\gamma(r); p)
\end{align*}
on $[0,\pi]$. Using the spherical volume element $d_{\mathbb{S}^{d}}\Omega(q)$ at $q=\gamma(r)$ we note
\begin{align*}
  1 = \int g(q;p) d_{\mathbb{S}^{d}}\Omega(q) = \int \frac{h(r;p)}{\text{vol}_{\mathbb{S}^{d-1}}} d_{\mathbb{S}^{d}}\Omega(q) = \int_0^\pi \limits h(r;p) \sin^{d-1}(r) dr \,
\end{align*}
which means that $h$ is a marginal distribution with respect to the measure
\begin{align*}
  d \mu_h(r) = \sin^{d-1}(r) dr \, .
\end{align*}
The marginal distribution with respect to the Lebesgue measure on $[0,\pi]$ is defined as
\begin{align*}
  f(r;p) := \sin^{d-1}(r) h(r;p) \, , \quad \int_0^\pi \limits f(r;p) dr = 1
\end{align*}

For the following, consider the so-called folded normal distribution
\begin{align*}
  \mathcal{F}(r; \rho, \sigma) :=& \frac{1}{\sqrt{2\pi} \sigma} \left( \exp\left(- \frac{(r-\rho\sigma)^2}{2 \sigma^2} \right) + \exp\left(- \frac{(r+\rho\sigma)^2}{2 \sigma^2}\right) \right) \\
  =& \frac{2}{\sqrt{2\pi} \sigma} \exp\left(- \frac{r^2}{2 \sigma^2} - \frac{\rho^2}{2} \right) \cosh\left(\frac{r\rho}{\sigma}\right), \,r\geq 0\, .
\end{align*}

For  $\rho \rar \infty$ this tends to a usual normal distribution centered at $\rho\sigma$, while it becomes a halved normal distribution for $\rho \rar 0$. Visualizing as a surface of revolution over $\mathbb{R}^2$, in polar coordinates $(r,\vartheta)\mapsto \mathcal{F}(r; \rho, \sigma)\,\frac{1}{2\pi}$, the former case yields a ring while the latter case yields a symmetric Gaussian distribution. Due to its smoothness it is a good candidate for a test distribution to distinguish concentrated clusters from ring shapes.

With the above marginals we therefore define
\begin{align*}
  h(r;p, \rho, \sigma) = \frac{\sqrt{2\pi} \sigma}{\mathcal{C}(\rho, \sigma)} \mathcal{F}(r; \rho, \sigma),\quad f(r;p,\rho,\sigma) = \frac{\sqrt{2\pi} \sigma}{\mathcal{C}(\rho, \sigma)} \sin^{d-1}(r) \mathcal{F}(r; \rho, \sigma) \, ,
\end{align*}
whose normalization $\mathcal{C}(\rho, \sigma)$ can be easily determined numerically. We can determine the MLEs for $\rho$ and $\sigma$ using standard numerical optimization. Although a numerical integral has to be calculated for normalization in each optimization step, the optimizations usually converge very quickly. If $\rho_{\text{MLE}} < 1$, the distribution has its maximum at $r=0$ and the small subsphere hypothesis can be readily rejected in favor of a great subsphere fit. 

If $\rho_{\text{MLE}} > 1$ we apply a likelihood ratio test. For a fixed $p$, given as the center of the best fit small subsphere, $q_i$ the data and $r_i = d(p,q_i)$ the spherical distances,
\begin{align*}
  \ell(\rho, \sigma | \{r_i\}_{i=1}^n) =& - n \ln \mathcal{C}(\rho, \sigma) + (d-1) \sum_{i=1}^n \limits \ln\sin(r_i)\\
  &- \frac{n\rho^2}{2} + n \ln(2) + \sum_{i=1}^n \limits \left( - \frac{r_i^2}{2 \sigma^2} + \ln\cosh\left(\frac{r_i \rho}{\sigma}\right)\right)
\end{align*}
is the log-likelihood for $f(r;p,\rho,\sigma)$ given a data set $\{r_i\}$. As null hypothesis we assume $\rho=1$, which means that the data form a dense cluster. The alternative hypothesis $\rho > 1$ means that the data are better approximated by a small circle. Let
\begin{align*}
  \lambda (\{r_i\}_{i=1}^n) =& 2 \sup\{ \ell(\rho, \sigma | \{r_i\}_{i=1}^n) : \rho \in (1, \infty), \, \sigma \in \mathbb{R}^+ \}\\
  &- 2 \sup\{ \ell(\rho, \sigma | \{r_i\}_{i=1}^n) : \rho = 1, \, \sigma \in \mathbb{R}^+ \}
\end{align*}
be the usual test statistic for a likelihood ratio test. Then, due to Wilks' theorem, see \cite{Wilks1938} and \cite[Chapter~16]{vanderVaart1998}, $\lambda$ is asymptotically $\chi^2_1$ distributed for $n \rar \infty$. The null hypothesis is rejected with $95 \%$ confidence if $\lambda > \chi^2_{1,95}$ in which case we keep the fitted small subsphere. If the null hypothesis is not rejected, we perform a great subsphere fit.

\begin{figure}[ht!]
   \centering
   \subcaptionbox{\it The probability distribution $h$ for any $d$}[0.45\textwidth]{\includegraphics[width=0.45\textwidth]{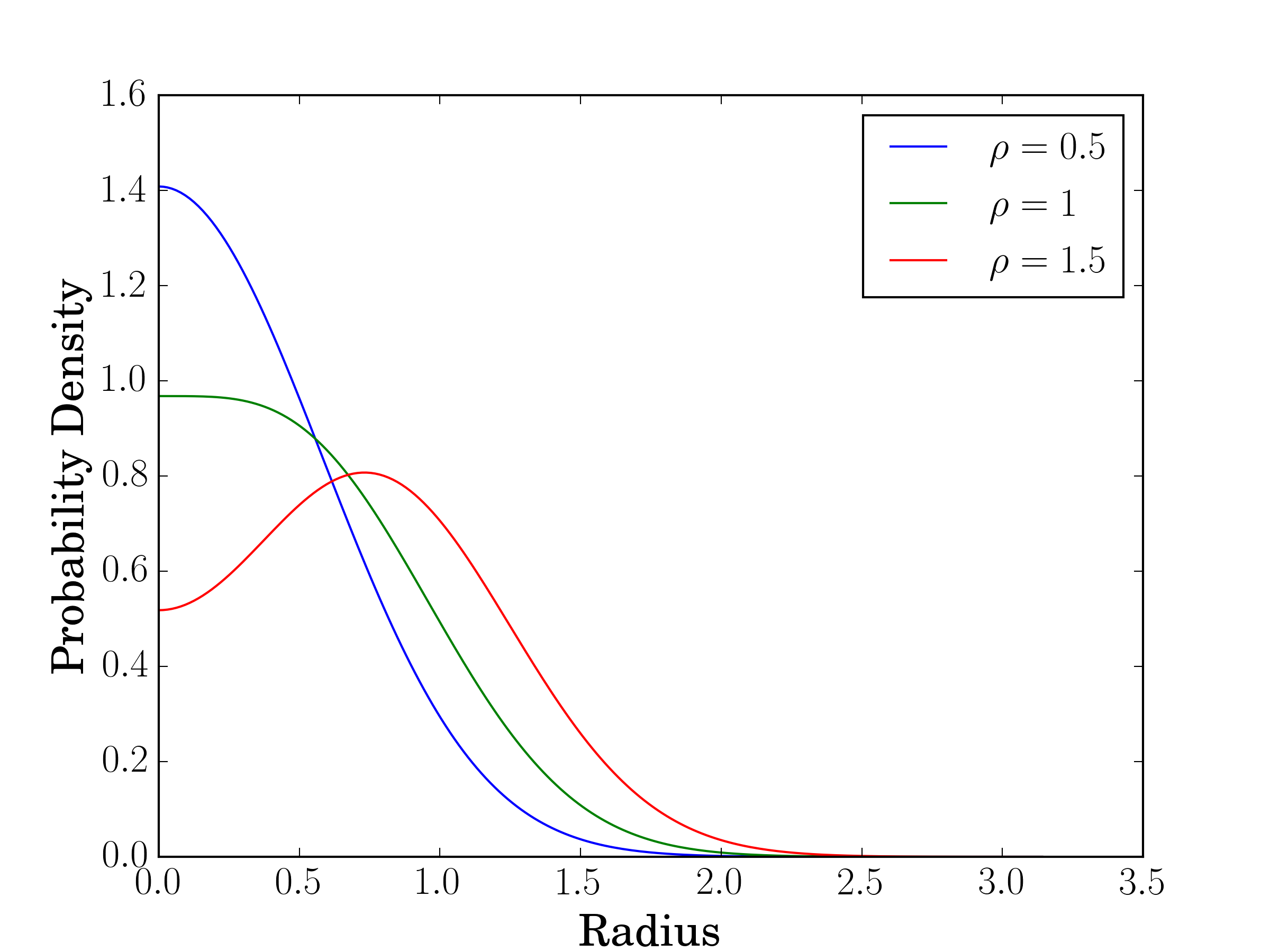}}
   \hspace*{0.05\textwidth}
   \subcaptionbox{\it The probability distribution $h^*$ for $d=2$}[0.45\textwidth]{\includegraphics[width=0.45\textwidth]{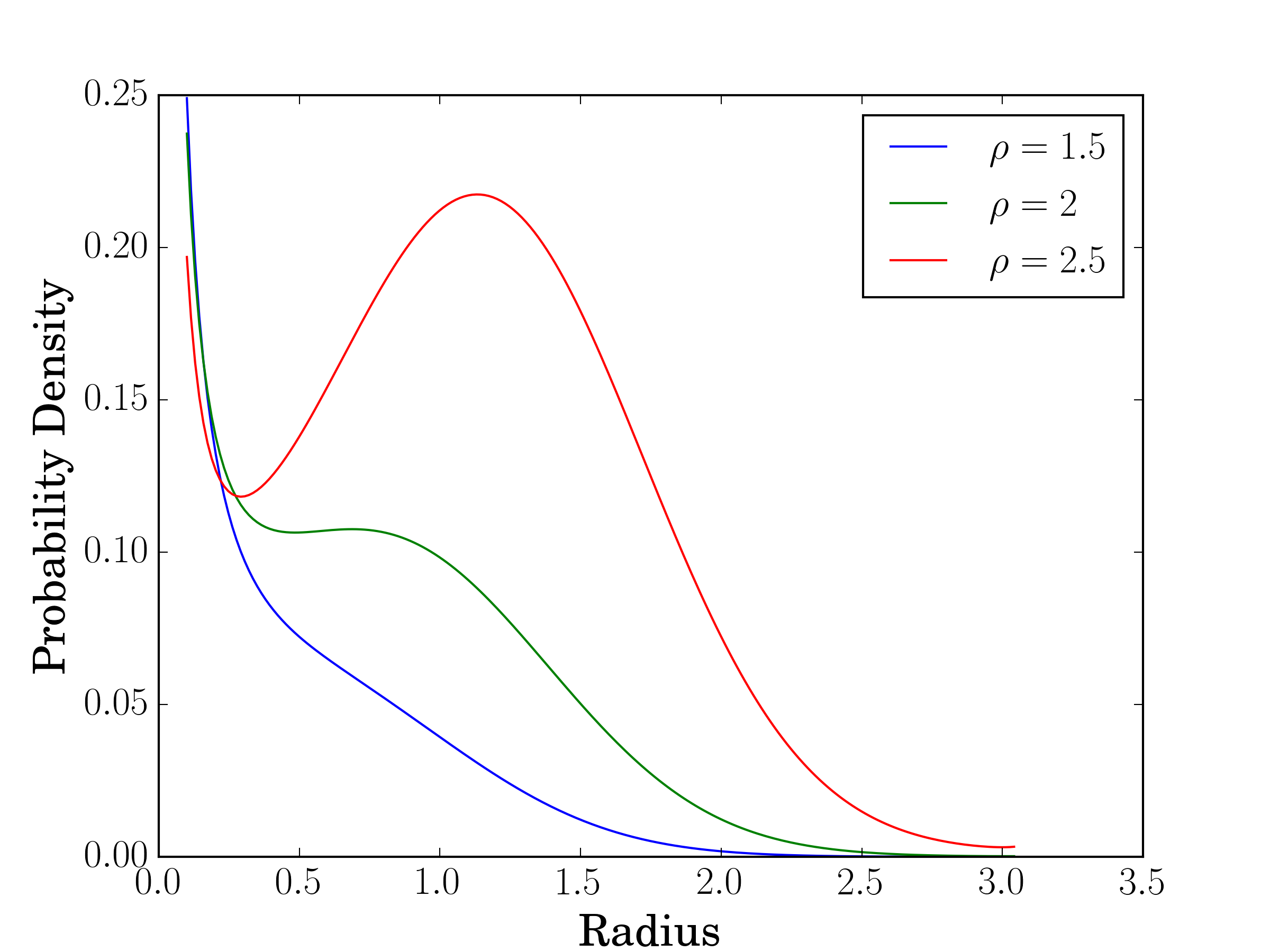}}
   \caption{\it The probability densities for $\sigma = 0.5$ along a geodesic in $\mathbb S^{d}$ for the distribution $h$ used here and $h^*$ used by \cite{Jung2011}. Displaying a value for $\rho$ below the respective threshold, at the threshold and above the threshold -- which is $1$ in our approach and $2$ in case of \cite{Jung2011}.}
   \label{prob_dens}
\end{figure}

The functions $f$ and $h$ are briefly discussed in \cite{Jung2011} but they use the folded normal distribution defined on $\mathbb{R}^+$ for $f$, i.~e.
\begin{align*}
  f^*(r;p, \rho, \sigma) = \mathcal{F}(r; \rho, \sigma), \quad h^*(r; p, \rho, \sigma) = \frac{1}{\sin^{d-1}(r)} \mathcal{F}(r; \rho, \sigma)\, ,
\end{align*}
leading to a singularity of the probability density $h^*$ and thus of $g$ at $p$. The small circle is accepted, if the probability distribution exhibits a ring shaped local maximum, which is the case for $\rho > 2$. A singularity at $p$, however, is an undesirable feature of the distribution $h^*$ (resulting in a frequent rejection of a projected Gaussian in the tangent space as null hypothesis as seen in Table \ref{test_comparison} in the original article), which we have avoided as above. The functions $h(r)$ for our distribution and $h^*(r)$ for the one used by \cite{Jung2011} are illustrated in Figure \ref{prob_dens}.

In \cite{Jung2012} among others, null hypothesis and alternative are both modeled via von Mises-Fisher distributions and a student $t$-like test statistic of distances to the estimated center point is used. However, the von Mises-Fischer distribution has a heavier tail and thus a higher standard deviation than the truncated Gaussians used here. In consequence, the test statistic is considerably larger for our more concentrated clusters and thus the null hypothesis is frequently rejected, especially for large sample sizes (see Table \ref{test_comparison} in the original article). 

\subsection{Single Linkage Branch Cutting Algorithm}

Let $P$ with $|P| =: n$ be the set of data points to cluster, $m$ a lower bound for the cluster size and $d_{\text{max}}$ a maximal outlier distance. Define the minimal cluster size $S_P = \sqrt{n + m^2}$. Then, we iteratively first store outliers to a list $R$ and cluster the rest of the data to a list $C$ as follows.
\begin{enumerate}
 \item Start with empty lists $R$ and $C$.
  \item Compute the cluster tree of $P$.
  \item Perform a branch cut at distance $d_{\text{max}}$, i.~e. removing all nodes with values above $d_{\text{max}}$. For all nodes with less than $m$ points add their points to the list $R$ and remove them from $P$.
  \item Compute $S_P$ and the cluster tree for $P$. 
  \item Perform the branch cut,
  \begin{enumerate}
    \item Starting from the root, follow the branch containing more points at each fork.
    \item If the smaller branch at a fork has more than $S_P$ points, store it to a list $L$.
    \item At the last fork, where the smaller branch has size at least $S_P$ store also the larger branch to $L$.
    \item For the largest cluster in $L$ remove its points from $P$ and store the cluster to a list $C$. If $L$ is empty, add all remaining points to $C$ as one cluster.
  \end{enumerate}
  \item If points remain, go to 2.
  \item Return the list of clusters $C$ and the outliers $R$.
\end{enumerate}

For our analysis we chose $m=15$ and $d_{\text{max}} = 50^\circ$.

\subsection{Outliers in the Small Data Set} 

\begin{figure}[ht!]
  \centering
  \subcaptionbox{\it 2D approximation, SI\label{geoPCA_outliersa}}[0.45\textwidth]{\includegraphics[width=0.45\textwidth, clip=true, trim=0.5cm 0.2cm 0.5cm 0.2cm]{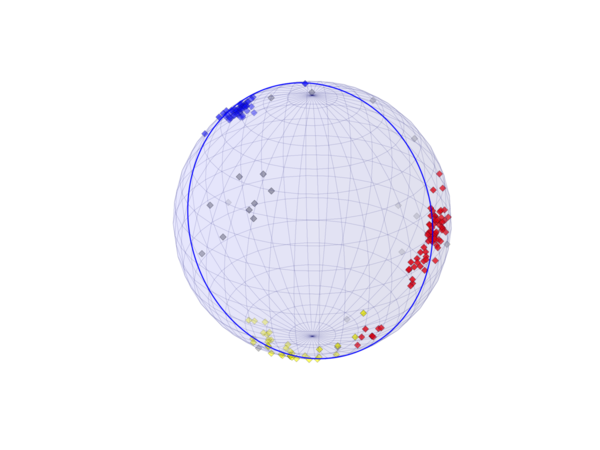}}
  \hspace*{0.05\textwidth}
  \subcaptionbox{\it 2D approximation, SO\label{geoPCA_outliersb}}[0.45\textwidth]{\includegraphics[width=0.45\textwidth, clip=true, trim=0.5cm 0.2cm 0.5cm 0.2cm]{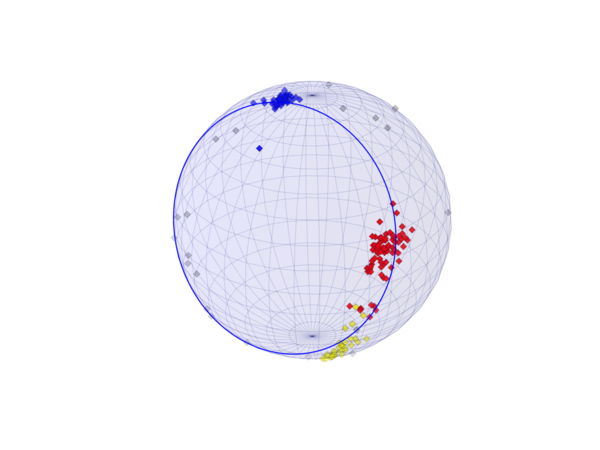}}
  \caption{\it Two dimensional T-PCA approximation of the pre-clusters found in the small RNA data set with SI (\ref{geoPCA_outliersb}) and SO (\ref{geoPCA_outliersb}) ordering. Red, blue and yellow represent the same clusters as in Figure \ref{geoPCA_labels}, gray points have been classified as outliers.}
  \label{geoPCA_outliers}
\end{figure}

The small data set has been devised by \cite{Sargsyan2012} to feature three clusters. These are found by T-PCA. Additionally including preclustering in our method, outliers are found whose small scale geometry is very different from the majority of the residues of any of the three clusters. Clusters and outliers are depicted in Figure \ref{geoPCA_outliers}.

\FloatBarrier

\subsection{List of All Found Clusters}

\begin{table}[!ht]
  \caption{\textit{Nested means of the $22$ clusters found by T-PCA.}}
  \begin{center}
    \begin{tabular}{|r|r|r||*{7}{r|}}
      \hline
      &&& \multicolumn{7}{|c|}{\rule{0pt}{2.6ex} nested mean} \\ \hline \vspace*{0.5\baselineskip}
      cluster \# & \# points & method & \multicolumn{1}{|c|}{$\alpha$} & \multicolumn{1}{|c|}{$\beta$} & \multicolumn{1}{|c|}{$\gamma$} & \multicolumn{1}{|c|}{$\delta$} & \multicolumn{1}{|c|}{$\epsilon$} & \multicolumn{1}{|c|}{$\zeta$} & \multicolumn{1}{|c|}{$\chi$} \\ \hline \vspace*{0.5\baselineskip}
      1  & 4921 & SO MC & 297.95 & 173.73 &  52.47 &  81.02 & 202.96 & 291.83 & 197.96\\ \vspace*{0.5\baselineskip}
      2  & 477  & SO MC & 162.62 & 203.96 & 171.62 &  83.72 & 232.57 & 280.81 & 181.95\\ \vspace*{0.5\baselineskip}
      3  & 232  & SO GC & 306.29 & 128.46 &  53.09 &  84.23 & 205.07 & 293.40 & 225.13\\ \vspace*{0.5\baselineskip}
      4  & 211  & SO MC & 157.78 & 182.50 &  51.04 &  84.70 & 212.65 & 292.49 & 195.22\\ \vspace*{0.5\baselineskip}
      5  & 145  & SO GC & 293.52 & 173.23 &  53.80 &  82.88 & 208.23 &  71.47 & 207.56\\ \vspace*{0.5\baselineskip}
      6  & 139  & SI GC &  56.76 & 160.75 &  51.56 &  83.65 & 216.26 & 289.84 & 189.47\\ \vspace*{0.5\baselineskip}
      7  & 137  & SO GC & 304.39 & 162.78 &  59.02 & 146.49 & 228.62 & 157.06 & 243.20\\ \vspace*{0.5\baselineskip}
      8  & 134  & SI GC & 294.62 & 173.66 &  53.57 &  83.41 & 227.20 & 204.37 & 204.05\\ \vspace*{0.5\baselineskip}
      9  & 125  & SO GC & 304.31 & 164.04 &  45.01 & 145.04 & 247.30 &  76.24 & 227.76\\ \vspace*{0.5\baselineskip}
      10 & 122  & SO GC & 210.51 & 115.51 & 160.16 &  85.67 & 225.35 & 280.39 & 184.57\\ \vspace*{0.5\baselineskip}
      11 & 107  & SO GC &  99.00 & 177.58 &  57.58 & 144.18 & 268.68 & 319.98 & 237.95\\ \vspace*{0.5\baselineskip}
      12 & 85   & SO GC & 303.40 & 165.94 &  59.13 & 142.65 & 260.53 & 292.92 & 237.45\\ \vspace*{0.5\baselineskip}
      13 & 84   & SI MC & 146.77 & 221.93 & 170.04 &  80.20 & 245.21 & 223.11 & 189.16\\ \vspace*{0.5\baselineskip}
      14 & 78   & SI MC &  79.50 & 188.62 & 182.21 &  84.75 & 214.53 & 293.24 & 201.44\\ \vspace*{0.5\baselineskip}
      15 & 60   & SO MC & 152.44 & 136.18 &  51.17 & 146.56 & 277.10 & 116.79 & 226.51\\ \vspace*{0.5\baselineskip}
      16 & 60   & SI GC & 280.79 & 199.65 & 174.56 &  90.66 & 217.55 & 271.56 & 210.26\\ \vspace*{0.5\baselineskip}
      17 & 59   & SO GC &  67.15 & 180.96 &  52.17 & 149.34 & 246.12 &  62.31 & 249.72\\ \vspace*{0.5\baselineskip}
      18 & 55   & SI MC &  50.61 & 184.82 & 286.25 &  97.58 & 205.34 & 306.97 & 202.51\\ \vspace*{0.5\baselineskip}
      19 & 52   & SO GC & 291.86 & 184.92 &  53.69 &  95.52 &  30.43 & 166.71 & 230.74\\ \vspace*{0.5\baselineskip}
      20 & 46   & SO MC & 254.42 & 199.20 &  80.36 &  97.36 & 302.41 & 233.82 & 219.45\\ \vspace*{0.5\baselineskip}
      21 & 33   & SO GC & 278.08 & 253.81 & 280.18 &  86.93 & 203.81 & 294.46 & 193.98\\ \vspace*{0.5\baselineskip}
      22 & 28   & SI MC & 290.03 & 199.31 &  55.09 &  87.25 &  71.04 & 283.36 & 226.96\\ \hline
    \end{tabular}
  \end{center}
  \label{table_all_clusters}
\end{table}

\begin{table}[!ht]
  \caption{\textit{Residual variances of the projections of the $22$ clusters found by T-PCA for all correspondingly dimensional subspheres.}}
  \begin{center}
    \begin{tabular}{|r|*{7}{r|}}
      \hline
      cluster \# & \multicolumn{1}{|c|}{0~D} & \multicolumn{1}{|c|}{1~D} & \multicolumn{1}{|c|}{2~D} & \multicolumn{1}{|c|}{3~D} & \multicolumn{1}{|c|}{4~D} & \multicolumn{1}{|c|}{5~D} & \multicolumn{1}{|c|}{6~D} \\ \hline \vspace*{0.5\baselineskip}
      1  &   653.67 &  394.26 &  277.06 &  176.02 &  93.88 &  38.47 &  24.38\\ \vspace*{0.5\baselineskip}
      2  &  2002.29 &  717.43 &  369.59 &  225.49 & 114.48 &  55.06 &  24.84\\ \vspace*{0.5\baselineskip}
      3  &  7548.59 & 2794.06 & 1647.43 & 1089.96 & 770.95 & 283.91 &  39.87\\ \vspace*{0.5\baselineskip}
      4  &  2540.55 &  649.73 &  334.17 &  213.59 & 118.32 &  60.94 &  33.17\\ \vspace*{0.5\baselineskip}
      5  &  3563.49 & 1096.97 &  568.65 &  342.73 & 221.13 & 134.87 &  66.42\\ \vspace*{0.5\baselineskip}
      6  &  4361.17 & 1436.75 &  846.12 &  598.74 & 302.46 & 112.93 &  16.78\\ \vspace*{0.5\baselineskip}
      7  &  3370.20 & 1279.69 &  735.58 &  522.58 & 267.93 & 151.08 &  69.19\\ \vspace*{0.5\baselineskip}
      8  &  1340.83 &  525.74 &  212.90 &  110.07 &  53.82 &  18.01 &   5.76\\ \vspace*{0.5\baselineskip}
      9  &  3513.08 & 1474.97 &  945.99 &  522.21 & 303.78 & 138.61 &  38.67\\ \vspace*{0.5\baselineskip}
      10 &  4377.74 & 1273.44 &  762.57 &  459.16 & 261.26 &  97.23 &  53.49\\ \vspace*{0.5\baselineskip}
      11 &  7065.66 & 1510.25 &  942.27 &  641.22 & 377.90 & 126.61 &  53.06\\ \vspace*{0.5\baselineskip}
      12 &  1977.41 &  919.47 &  580.16 &  367.15 & 232.95 & 113.00 &  46.77\\ \vspace*{0.5\baselineskip}
      13 &  7622.07 & 2997.64 & 1683.83 & 1098.01 & 613.00 & 236.14 & 106.52\\ \vspace*{0.5\baselineskip}
      14 &  4570.60 & 2194.72 & 1365.03 &  664.37 & 398.73 & 176.38 &  18.33\\ \vspace*{0.5\baselineskip}
      15 &  4379.46 & 1273.97 &  761.20 &  505.75 & 204.07 &  99.97 &  30.46\\ \vspace*{0.5\baselineskip}
      16 & 24749.45 & 2334.71 & 1455.38 &  809.90 & 492.65 & 242.71 &  56.44\\ \vspace*{0.5\baselineskip}
      17 &  7771.96 & 2046.17 & 1304.37 &  807.84 & 467.27 & 202.97 &  66.31\\ \vspace*{0.5\baselineskip}
      18 &  2357.09 & 1325.03 &  766.96 &  435.02 & 283.81 & 161.77 &  49.54\\ \vspace*{0.5\baselineskip}
      19 &  9360.10 & 1660.47 & 1056.75 &  668.63 & 368.23 & 173.33 &  66.23\\ \vspace*{0.5\baselineskip}
      20 & 11420.75 & 3021.12 & 1762.57 &  952.27 & 420.14 & 218.83 &  97.05\\ \vspace*{0.5\baselineskip}
      21 &  4395.15 & 1127.39 &  479.15 &  232.76 & 146.39 & 108.96 &  44.76\\ \vspace*{0.5\baselineskip}
      22 &  2926.13 & 1538.61 &  986.94 &  613.66 & 288.46 & 157.12 &  70.01\\ \hline
    \end{tabular}
    \end{center}
  \label{table_all_clusters2}
\end{table}

\begin{table}[!ht]
  \caption{\textit{The $15$ pre-clusters and their decomposition into the $22$ clusters found by mode hunting.}}
  \begin{center}
    \begin{tabular}{|r|r|r|}
      \hline
      pre-cluster \# & \# points & cluster \#s \\ \hline \vspace*{0.5\baselineskip}
      1  & 5055 & 1, 8\\ \vspace*{0.5\baselineskip}
      2  & 492  & 5, 7, 9, 12\\ \vspace*{0.5\baselineskip}
      3  & 477  & 2\\ \vspace*{0.5\baselineskip}
      4  & 232  & 3\\ \vspace*{0.5\baselineskip}
      5  & 226  & 11, 15, 17\\ \vspace*{0.5\baselineskip}
      6  & 211  & 4\\ \vspace*{0.5\baselineskip}
      7  & 139  & 6\\ \vspace*{0.5\baselineskip}
      8  & 133  & 14, 18\\ \vspace*{0.5\baselineskip}
      9  & 122  & 10\\ \vspace*{0.5\baselineskip}
      10 & 84   & 13\\ \vspace*{0.5\baselineskip}
      11 & 60   & 16\\ \vspace*{0.5\baselineskip}
      12 & 52   & 19\\ \vspace*{0.5\baselineskip}
      13 & 46   & 20\\ \vspace*{0.5\baselineskip}
      14 & 32   & 21\\ \vspace*{0.5\baselineskip}
      15 & 28   & 22\\ \hline
    \end{tabular}
  \end{center}
  \label{table_all_clusters3}
\end{table}

\FloatBarrier

\begin{figure}[ht!]
  \centering
  \subcaptionbox{\it Cluster 11}[0.45\textwidth]{\includegraphics[width=0.45\textwidth, clip=true, trim=0.5cm 0.2cm 0.5cm 0.2cm]{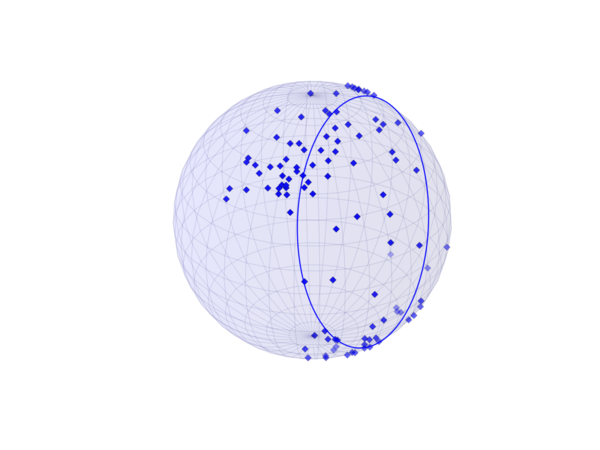}}
  \hspace*{0.05\textwidth}
  \subcaptionbox{\it Cluster 16\label{divisible_clustersb}}[0.45\textwidth]{\includegraphics[width=0.45\textwidth, clip=true, trim=0.5cm 0.2cm 0.5cm 0.2cm]{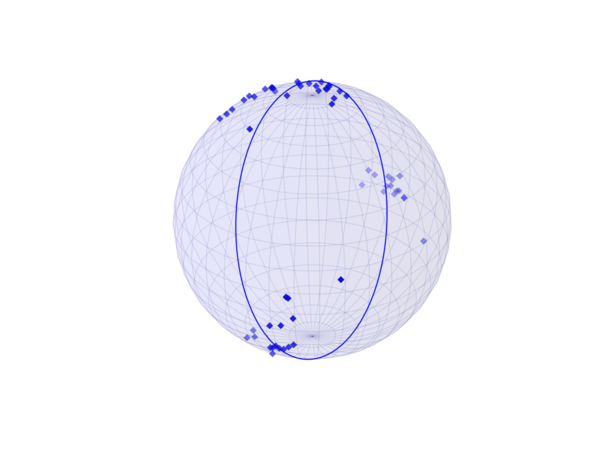}}
  \caption{\it Two dimensional T-PCA approximations of two clusters which appear to be composed of several distinct clusters.}
  \label{divisible_clusters}
\end{figure}

Visual inspection of the one- or two-dimensional projections shows distinct subgroups for some clusters, see Figure \ref{divisible_clusters}. These subgroups have not been found by T-PCA due to the $95\%$ confidence bound for mode hunting; in particular in Figure \ref{divisible_clustersb} the number of points is too small to attain significance.

\FloatBarrier

\bibliographystyle{../bib/apalike}
\bibliography{../bib/stat,../bib/shapstat,../bib/bio}

\end{document}